\documentclass[twocolumn]{autart}    

\usepackage{graphicx}          
                               
\usepackage{epsfig} 
\usepackage{newtxtext}
\usepackage{times} 
\usepackage{amsmath} 
\usepackage{amssymb}  
\usepackage{mathtools}
\usepackage{verbatim}
\usepackage{cases}
\usepackage{caption}
\usepackage{subcaption}
\usepackage{algorithm}
\usepackage[noend]{algpseudocode}
\usepackage{etoolbox}
\usepackage{multirow}
\usepackage{color}
\usepackage{harvard}

\makeatletter
\DeclareRobustCommand{\qed}{%
  \ifmmode 
  \else \leavevmode\unskip\penalty9999 \hbox{}\nobreak\hfill
  \fi
  \quad\hbox{\qedsymbol}}
\newcommand{\openbox}{\leavevmode
  \hbox to.77778em{%
  \hfil\vrule
  \vbox to.675em{\hrule width.6em\vfil\hrule}%
  \vrule\hfil}}
\newcommand{\qedsymbol}{\openbox}
\newenvironment{proof}[1][\proofname]{\par
  \normalfont
  \topsep6\p@\@plus6\p@ \trivlist
  \item[\hskip\labelsep\itshape
    #1.]\ignorespaces
}{%
  \qed\endtrivlist
}
\newcommand{\proofname}{Proof}

\makeatother

\newcommand{\lang}{\mathcal{L}}

\usepackage{enumitem}
\begin{document}
\graphicspath{{Fig/}}

\begin{frontmatter}

\title{Synthesis of Sensor Deception Attacks at the Supervisory Layer of Cyber-Physical Systems} 

\thanks[footnoteinfo]{This work was supported in part by the US National Science Foundation under grants CNS-1421122, CNS-1446298 and CNS-1738103.}

\author[Ann Arbor]{R\^omulo Meira-G\'oes}\ead{romulo@umich.edu},    
\author[Pit]{Eunsuk Kang}\ead{eskang@cmu.edu},    
\author[Toronto]{Raymond H. Kwong}\ead{kwong@control.utoronto.ca},    
\author[Ann Arbor]{St\'ephane Lafortune}\ead{stephane@umich.edu}               

\address[Ann Arbor]{Department of Electrical Engineering and Computer Science, University of Michigan, Ann Arbor, MI 48109, USA}
\address[Pit]{School of Computer Science, Carnegie Mellon University, Pittsburgh, PA 15213, USA} 
\address[Toronto]{Department of Electrical and Computer Engineering, University of Toronto, Toronto, ON M5S 3G4, CA}  

\begin{keyword}                           
Discrete Event Systems; Supervisory Control; Cyber-Physical Systems; Cyber-Security; Deception Attacks.               
\end{keyword}                             

\begin{abstract}                          
We study the security of Cyber-Physical Systems (CPS) in the context of the supervisory control layer. 
Specifically, we propose a general model of a CPS attacker in the framework of discrete event systems and investigate the problem of synthesizing an attack strategy for a given feedback control system. 
Our model captures a class of \textit{deception attacks}, where the attacker has the ability to hijack a subset of sensor readings and mislead the supervisor, with the goal of inducing the system into an undesirable state. 
We utilize a game-like discrete transition structure, called \textit{Insertion-Deletion Attack structure} (IDA), to capture the interaction between the supervisor and the environment (which includes the system and the attacker). 
We show how to use IDAs to synthesize three different types of successful stealthy attacks, i.e., attacks that avoid detection from the supervisor and cause damage to the system.
\end{abstract}

\end{frontmatter}

\section{Introduction}

Cyber-Physical Systems (CPS) are characterized by the interaction of computational entities with physical processes. 
These systems are found in a broad spectrum of safety-critical applications, such as smart grids, process control systems, autonomous vehicles, medical devices, etc. 
In such applications, some undesired behavior could cause damage to the physical system itself or to people relying on the system. 
Undesired behavior of CPS may be caused by faulty behavior or by external attacks on the system.
In fact, some attacks on CPS have already been reported, see, e.g., \cite{Kerns:2014,Checkoway:2011}, including the well-known StuxNet attack \cite{Farwell:2011}. 

For this reason, cyber-security of CPS became a subject of increasing attention in the control community, see, e.g., \cite{Weerakkody:2019}. 
The work in \cite{Cardenas:2008,Teixeira:2012} have classified different types of cyber attacks on control systems.
These cyber attacks have assumed attackers that have some prior knowledge of the CPS.
Our paper focuses on security issues that arise in the feedback control of CPS for a specific class of these cyber attacks called \emph{sensor deception attacks}.
In this class of attacks, the sensor readings are hijacked by an attacker that is assumed to have some prior knowledge of the control system.
More specifically, we are concerned with the problem of synthesizing a \textit{sensor deception attack} strategy at the supervisory layer of a given CPS \cite{Amin:2013}, an attack strategy that manipulates the sensor measurements received by the controller/supervisor\footnote{Since our focus is the supervisory layer of the control system, we will use the term \emph{supervisor} in the remainder of the paper.} in order to achieve the attacker's goals. 
Three different types of attacks are investigated, where each attack has different capabilities regarding manipulating the sensor measurements that are sent to the supervisor.

Given that we are investigating cyber-attacks at the supervisory layer of a CPS, we use the formalism of Discrete Event Systems (DES) to model both the behavior of the attacker as well as the behavior of CPS itself. 
In other words, we assume that a discrete abstraction of the underlying CPS has already been performed.
This allows us to leverage the concepts and techniques of the theory of supervisory control of DES.
Several recent works have adopted similar approaches to study cyber-security issues in CPS;
see, e.g., \cite{Carvalho:2018,Paoli:2011,Thorsley:2006,Wakaiki:2017}.

Previous works such as \cite{Carvalho:2018,Paoli:2011,Thorsley:2006} on intrusion detection and prevention of cyber-attacks using discrete event models were focused on modeling the attacker as faulty behavior.
Their corresponding methodologies relied on fault diagnosis techniques.

Recently, \cite{Su:2018} proposed a framework similar to the one adopted in our paper, where they formulated a model of bounded sensor deception attacks.  
Our approach is more general than the one in \cite{Su:2018}, since we do not impose a \emph{normality} condition to create an attack strategy; this condition is imposed to obtain the so-called \emph{supremal controllable and normal language} under the attack model. 
In addition to bounded sensor deception attacks, we consider two other attack models.
An additional difference between this paper and the approach in \cite{Su:2018} is the way the dynamical interaction between the attacker and the supervisor is captured. This will be revisited in Section~\ref{sec:deception}.

In \cite{Wakaiki:2017}, the authors presented a study of supervisory control of DES under attacks. 
They introduced a new notion of \emph{observability} that captures the presence of an attacker. 
However, their study is focused on the supervisor's viewpoint and they do not develop a methodology to design attack strategies. 
They assume that the attack model is given and they develop their results based on that assumption. 
In that sense,  the work \cite{Wakaiki:2017} is closer to robust supervisory control and it is complementary to our work. 

Several prior works considered robust supervisory control under different notions of robustness \cite{Alves:2014,Lin:2014,Rohloff:2012,Xu:2009,Yin:2016c}, but they did not study robustness against attacks. 
In the cyber-security literature, some works have been carried out in the context of discrete event models, especially regarding opacity and privacy or secrecy properties \cite{Cassez:2012,Lin:2011,Saboori:2007,Wu:2018,Meira-Goes:2018}. 
These works are concerned with studying information release properties of the system, and they do not address the impact of an intruder over the physical parts of the system. 

Our approach is based on a general model of sensor deception attacks at the supervisory layer of the control system of the CPS.
In that context, we investigate the problem of synthesizing \emph{successful stealthy sensor deception attacks}. 
We make the following assumptions about the attacker: (i) it has knowledge of both the system and its supervisor; and (ii) it has the ability to alter the sensor information that is received by the supervisor.
The goal of the attacker is to induce the supervisor into allowing the system to reach a pre-specified unsafe state, thereby causing damage to the system.
Throughout the paper, we discuss the reasoning and the impact of these assumptions.

In this paper, we study three specific types of attacks that are based on the interaction between the attacker and the controlled system.
The methodology developed to synthesize these attacks is inspired by the work in \cite{Wu:2018,Yin:2016b,Yin:2016a}.
As in these works, we employ a discrete structure to model the game-like interaction between the supervisor and the environment (system and attacker in this paper).
We call this structure an \textit{Insertion-Deletion Attack} structure (or IDA). 
By construction, an IDA embeds all desired scenarios where the attacker modifies some subset of the sensor events without being noticed by the supervisor. 
Once constructed according to the classes of attacks under consideration, an IDA serves as the basis for solving the synthesis problem. 
In fact, this game-theoretical approach provides a structure for each attack class that incorporates \emph{all successful stealthy attacks}.
Different stealthy attack strategies can be extracted from this structure.
This is a distinguishing feature of our work as compared to previous works mentioned above.
By providing a general synthesis framework, our goal is to allow CPS engineers to detect and address potential vulnerabilities in their control systems.

The remainder of this paper is organized as follow. 
Section~\ref{sec:modeling} introduces necessary background and some notations used throughout the paper. The attack model as well as the problem statement are formalized in Section~\ref{sec:deception}. 
Section~\ref{sec:IDA} describes the IDA structure and its properties. 
Section~\ref{sec:AIDA} introduces the AIDA (All Insertion-Deletion Attack structure) and provides a construction algorithm for it. 
Sections~\ref{sec:synthesis-inter} and \ref{sec:othersynthe} introduce for each attack type their stealthy IDA structure, together with a simple synthesis algorithm to extract an attack function. 
Lastly, Section \ref{sec:conclusion} presents concluding remarks.
Preliminary and partial versions of some of the results in this work  have appeared in \cite{Goes:2017}. 
All proofs are located in the Appendix. 

\section{Supervisory control system model}\label{sec:modeling} 

We assume that a given CPS has been abstracted as a discrete transition system that is modeled as a finite-state automaton. 
A finite-state automaton $G$ is defined as a tuple $G = (X,\Sigma,\delta,x_0)$, where:
$X$ is a finite set of states;
$\Sigma$ is a finite set of events;
$\delta: X\times \Sigma \rightarrow X$ is a partial transition function;
and
$x_0 \in X$ is the initial state.
%
The function $\delta$ is extended in the usual manner to domain $X\times\Sigma^*$. 
The language generated by $G$ is defined as $\mathcal{L}(G) = \{s \in \Sigma^*| \delta(x_0, s)!\}$, where ! means ``is defined''.

In addition, $\Gamma_G(S)$ is defined as the set of active events at the subset of states  $S \subseteq X$ of automaton $G$, given by:
\begin{align}
\Gamma_G(S) := \{e\in\Sigma|(\exists u \in S)\text{ s.t. } \delta(u,e)!\}
\end{align}
By a slight abuse of notation, we write $\Gamma_G(x) = \Gamma_G(\{x\})$ for $x\in X$.
 
Language $\mathcal{L}(G)$ is considered as the \emph{uncontrolled} system behavior, since it includes all possible executions of $G$. 
The limited actuation capabilities of $G$ are modeled by a partition in the event set $\Sigma = \Sigma_c\cup\Sigma_{uc}$, where $\Sigma_{uc}$ is the set of uncontrollable events and $\Sigma_c$ is the set of controllable events.

It is assumed that $G$ is controlled by a supervisor $S_P$ that dynamically enables and disables the controllable events such that it enforces some safety property on $G$.
In the notation of the theory of supervisory control of DES initiated in \cite{Ramadge:1987}, the resulting controlled behavior is a new DES denoted by $S_P /G$ with the closed-loop language $\mathcal{L}(S_P/G)$ defined in the usual manner \cite{Lafortune:2008}.
The set of admissible control decisions is defined as $\Gamma = \{\gamma\subseteq \Sigma\mid \Sigma_{uc} \subseteq \gamma\}$, where admissibility guarantees that a control decision never disables uncontrollable events. 

In addition, due to the limited sensing capabilities of $G$, the event set is also partitioned into $\Sigma = \Sigma_o\cup\Sigma_{uo}$, where $\Sigma_o$ is the set of observable events and $\Sigma_{uo}$ is the set of unobservable events. 
Based on this second partition, the \textit{projection} function $P_o:\Sigma^*\rightarrow\Sigma_o^*$ is defined as:
\begin{align}
P_o(\epsilon) = \epsilon \text{ and } P_o(se)=\begin{cases}P_o(s)e & \text{if }e\in\Sigma_o\\P_o(s) & \text{if }e\in\Sigma_{uo}\end{cases}
\end{align}
The inverse projection $P^{-1}_o:\Sigma_o^*\rightarrow2^{\Sigma^*}$ is defined as $P^{-1}_o(t) = \{s\in\Sigma^*|P(s) = t\}$.

Formally, a partial observation supervisor is a function $S_P: P_o(\mathcal{L}(G))\rightarrow\Gamma$. 
Without loss of generality, we assume that $S_P$ is realized (i.e., encoded) as a deterministic automaton 
$R = (Q,\Sigma,\mu,q_0)$,
such that, $\forall q\in Q$, if $e \in \Sigma_{uo}$ is an enabled unobservable event at state $q$ by $S_P$, then we define $\mu(q,e)=q$ as is customary in a supervisor realization (cf.\ \cite{Lafortune:2008}).
In this manner, given a string $s\in \mathcal{L}(G)$ the control decision is defined as $S_P(s) =\Gamma_R( \mu(q_0,s))$. 



\begin{figure}[thpb]
      \centering
      \begin{subfigure}[t]{.5\columnwidth}
  		\centering
  		\includegraphics[width=.8\columnwidth]{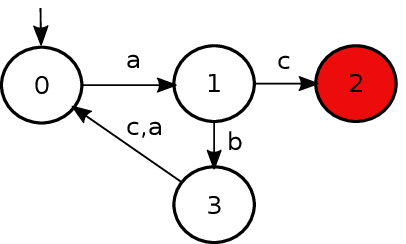}
  		\caption{$G$}
  		\label{fig:sys_G}
	\end{subfigure}%
	\begin{subfigure}[t]{.5\columnwidth}
  		\centering
  		\includegraphics[width=.55\columnwidth]{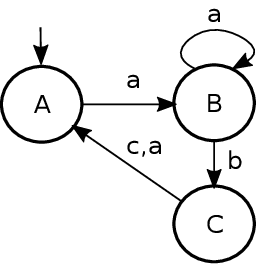}
  		\caption{Supervisor $R$}
  		\label{fig:sup_S}
		\end{subfigure}
	
	\caption{A system automaton along with its supervisor (Example~\ref{ex1})}
	\label{fig:ex1}
   \end{figure}

\begin{exmp}\label{ex1}
Consider the system $G$ represented in Fig.~\ref{fig:ex1}(\subref{fig:sys_G}). 
Let $\Sigma = \Sigma_o = \{a,b,c\}$ and $\Sigma_c = \{b,c\}$. 
Figure~\ref{fig:ex1}(\subref{fig:sup_S}) shows the realization $R$ of a supervisor $S_{P}$ that was designed for $G$. 
In this case, the language generated by $\mathcal{L}(S_{P}/G)$ guarantees that state 2 is unreachable in the controlled behavior. 
\end{exmp}

For convenience, we define two operators that are used in this paper together with some useful notation. 
The \emph{unobservable reach} of the subset of states $S \subseteq X$ under the subset of events $\gamma\subseteq\Sigma$ is given by:
\begin{align}
UR_\gamma(S) := &\{x\in X| (\exists u \in S)(\exists s \in (\Sigma_{uo}\cap\gamma)^*)\text{ s.t.}\nonumber \\&x = \delta(u,s)\}
\end{align}
The \emph{observable reach} (or next states) of the subset of states $S\subseteq X$ given the execution of the observable event $e\in\Sigma_o$ is defined as:
\begin{align}
NX_e(S) := \{x\in X | \exists u \in S \text{ s.t. } x = \delta(u,e)\}
\end{align}
For any string $s\in \Sigma^*$, let $s^i$ denote the prefix of $s$ with the first $i$ events, and let $e^i_s$ be the $i^{th}$ event of $s$, so that $s^i =e^1_s\ldots e^i_s$; by convention, $s^0  = \epsilon$.
We define $\bar{s}$ as the set of prefixes of string $s\in\Sigma^*$. 
Lastly, we denote by $\mathbb{N}$, $\mathbb{N}^{+}$, and $\mathbb{N}^{n} = \{0,\ldots,n\}$ the sets of natural numbers, positive natural numbers, and natural numbers bounded by $n$, respectively.
\section{The sensor deception attack problem}\label{sec:deception}
\subsection{The general attack model}
\begin{figure}[thpb]
      \centering
      \includegraphics[width=.65\columnwidth]{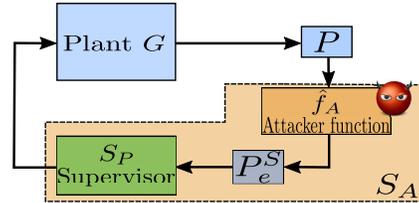}
      \caption{Model for Sensor Deception Attacks}
      \label{fig:model_cdc}
\end{figure}

We start by defining the model for sensor deception attacks, as illustrated in  Fig.~\ref{fig:model_cdc}. 
The attacker intervenes in the communication channels between the system's sensors and the supervisor.
We assume that the attacker observes all events in $\Sigma_o$ that are executed by the system. 
In addition, it has the ability to edit some of the sensor readings in these communication channels, by inserting fictitious events or deleting the legitimate events.
The subset of sensor readings that can be edited is defined as the \textit{compromised event set} and denoted by $\Sigma_a$.
For generality purposes, we assume that $\Sigma_a\subseteq\Sigma_o$.
To formally introduce the \emph{attack function} shown in Fig.~\ref{fig:model_cdc}, we first define two new sets of events. 
The sets $\Sigma_a^i = \{e_i| e \in \Sigma_a\}$ and $\Sigma_a^d = \{e_d| e \in \Sigma_a\}$ are defined as the sets of inserted and deleted events, respectively. 
These sets represent the actions of the attacker, and we use the subscripts to clearly distinguish them from events generated by  $G$.
Note that, $\Sigma_a^i$, $\Sigma_a^d$, and $\Sigma_a$ are disjoint.
For convenience, we define $\Sigma_a^e = \Sigma_a^i\cup\Sigma_a^d$ as the set of editable events. 
These events are used to identify the insertion or deletion of an event. 

We also define the projections $P^S_e$, $P^G_e$, and the mask $\mathcal{M}_e$ to analyze strings consisting of events in $\Sigma_a^e\cup\Sigma$. 
$P^S_e$ is a natural projection that treats editable events in the following manner: $P^S_e(e_i) = e$, $e_i\in\Sigma_a^i$; $P^S_e(e_d)=\epsilon$, $e_d \in \Sigma_a^d$, and $P^S_e(e) = e$, $e\in\Sigma$. 
We use the superscript $S$ because $P^S_e$ describes how the \emph{supervisor} observes the modified events.
Given an event, $P^S_e$ outputs the legitimate event if the event was inserted by the attacker, it outputs the empty string if the event was deleted by the attacker, and it outputs the legitimate event otherwise. 
In addition, we define $P^G_e$ to describe how the modified events interact with the system $G$: $P^G_e(e_i)=\epsilon$, $e_i\in\Sigma_a^i$; $P^G_e(e_d)=e$, $e_d \in \Sigma_a^d$, and $P^G_e(e) = e$, $e\in\Sigma$.
Finally, $\mathcal{M}_e$ is a mask that removes the subscript $\{i,d\}$ if it exists: $\mathcal{M}_e(e_i)= \mathcal{M}_e(e_d) = e$, if $e_i,e_d \in \Sigma_a^e$, and $\mathcal{M}_e(e) = e$, $e\in\Sigma$.

Formally, we model an attacker as a nondeterministic string edit function.
The nondeterminism model provides different class of attack strategies when compared to the deterministic model in \cite{Goes:2017}. 
\begin{defn}\label{def:attacker}
Given a system $G$ and a subset $\Sigma_a\subseteq\Sigma_o$, an attacker is defined as a (potentially partial) function $f_A:(\Sigma_o\cup\Sigma_a^e)^* \times (\Sigma_{o}\cup \{\epsilon\}) \rightarrow 2^{(\Sigma_o\cup\Sigma_a^e)^*}$ s.t.\ $f_A$ satisfies the following constraints:
\vspace*{-0.4cm}
\begin{enumerate}
\item $f_A(\epsilon,\epsilon) \subseteq{\Sigma_a^i}^*$ and $\forall s \in ((\Sigma_o\cup\Sigma_a^e)^*\backslash\{\epsilon\}):\ f_A(s,\epsilon)=\emptyset$;
\item $\forall s \in (\Sigma_o\cup\Sigma_a^e)^*$, $e\in\Sigma_o \backslash \Sigma_a $ : $f_A(s,e) \subseteq\ \{e\}{\Sigma_a^i}^*$;
\item $\forall s \in (\Sigma_o\cup\Sigma_a^e)^*$, $e\in\Sigma_a$: $f_A(s,e) \subseteq \ \{e,\ e_d\}{\Sigma_a^i}^*$.
\end{enumerate}
\end{defn}

The function $f_A$ captures a general model of sensor deception attack. 
Given the past \emph{edited} string $s$ and observing a new event $e$ executed by $G$, the attacker may choose to edit $e$ based on $\Sigma_a$ and replace $e$ by selecting an edited suffix from the set $f_A(s,e)$. 
The first case in the above definition gives an initial condition for an attack. 
The second case constrains the attacker from erasing $e$ when $e$ is outside of $\Sigma_a$. 
However, the attacker may insert an arbitrary string $t\in{\Sigma^i_a}^*$ after the occurrence of $e$. 
Lastly, the third case in Definition~\ref{def:attacker} is for $e\in\Sigma_a$; the attacker can edit the event to any string in the set $\{e,\ e_d\}{\Sigma_a^i}^*$.

As mentioned before, $f_A$ only defines the possible edited suffixes based on the last executed event and the edit history.
It is interesting to define a function that defines the possible edited strings based on the executed string.
Formally, the string-based edit (potentially partial) function $\hat{f}_A:\Sigma_o^* \rightarrow 2^{(\Sigma_o\cup\Sigma_a^e)^*}$ is defined as $\hat{f}_A(se)$ = $\{ ut \in (\Sigma_o\cup\Sigma_a^e)^*| u\in \hat{f}_A(s) \wedge t \in f_A(u,e)\}$ for any $s\in \Sigma_o^*$ and $e\in\Sigma_o$, and $\hat{f}_A(\epsilon)=f_A(\epsilon,\epsilon)$. 
The function $\hat{f}_A(s)$ returns the set of possible edited strings for a given string $s\in\Sigma_o^*$.  
Note that, in general, $\hat{f}_A$ is a partial function, and $\hat{f}_A(s)$ may only be defined for selected $s\in \Sigma_o^*$.

\emph{Remark:} We have assumed that the attacker has the same observable capabilities as the supervisor.
This assumption is accepted in papers, like ours, where the worst case analysis is performed.
The problem where the attacker and the supervisor has incomparable observation is known to be hard.
Nevertheless, in some cases the attacker might only have interception capabilities but full not transmission capabilities since it has not taken control of all the hardware.
For example, this assumption holds in wireless sensor networks.
\subsection{The controlled behavior under sensor deception attack}

\begin{figure}[thpb]
	\centering
	\begin{subfigure}[t]{.45\columnwidth}
		\centering
		\includegraphics[width=.8\columnwidth]{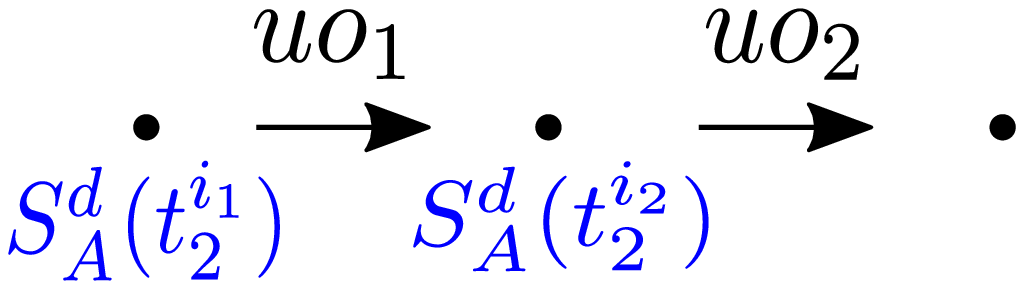}
		\caption{Only unobservable events}
		\label{fig:langunobs}
	\end{subfigure}%
	\begin{subfigure}[t]{.45\columnwidth}
		\centering
		\includegraphics[width=1\columnwidth]{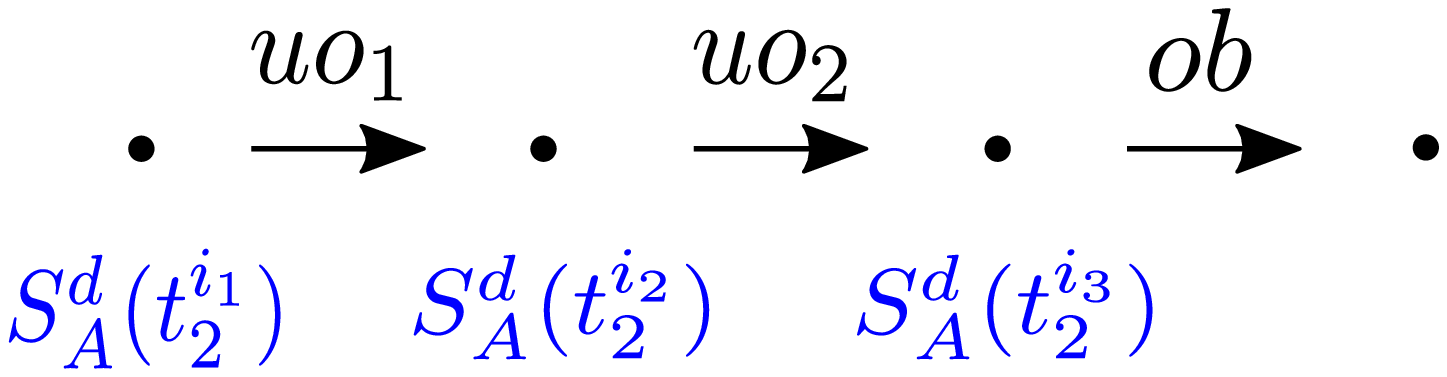}
		\caption{Last event is observable}
		\label{fig:langobs}
	\end{subfigure}
	\caption{Demonstration of Condition (2)}
	\label{fig:lang}
\end{figure}

The presence of the attacker induces a new controlled language that needs to be investigated.
More specifically, $S_P$, $\hat{f}_A$, and $P^S_e$ together effectively generate a new supervisor $S_A$ for system $G$, as depicted in Fig.~\ref{fig:model_cdc}.
Formally, we define $S_A:\Sigma_o^*\rightarrow 2^{\Gamma}$ as $S_A(s) = [S_P\circ P^S_e\circ \hat{f}_A(s)]$. Note that $\hat{f}_A(s)$ returns the set of modified strings and $S_P$ assigns a control decision to each projected modified string; $S_A$ returns the set containing all these control decisions.  
An equivalent definition is $S_A(s) = \{\gamma|\ \exists s_A\in\hat{f}_A(s)\text{ s.t. }\gamma = S_P(P_e^S(s_A))\}$.
Defining the supervised language based on nondeterministic control decisions is cumbersome and complicated \cite{Lin:2014}.
Nonetheless, we can avoid this difficulty by analyzing the language generated by the events in $\Sigma\cup\Sigma^a_e$.
For this reason, we define the function $S_A^d = S_P\circ P_e^S$ to be the deterministic part of $S_A$.
The function $S_A^d$ is used when the attacker has decided which modified string to send to the supervisor.
Based on $f_A$ and $S_A^d$ the language generated by $S_A/G$ is defined recursively as follows:
\begin{enumerate}
\item $\epsilon\in\mathcal{L}(S_A/G)$
\item $\big(t_1 \in \lang(G)\cap\Sigma_{uo}^*(\Sigma_o\cup\{\epsilon\})\big) \wedge \big(\exists t_2 \in f_A(\epsilon,\epsilon)$ and $\ i_1\leq i_2\leq \hdots \leq i_{|t_1|} \in \mathbb{N}^{|t_2|}$ s.t. $\forall j \in \mathbb{N}^{|t_1|}: e^j_{t_1} \in S^d_A(t_2^{i_j}) \big)\wedge \big(P_o(t_1)\neq \epsilon \Rightarrow i_{|t_1|} = |t_2|\big)\ \Leftrightarrow t_1 \in \lang(S_A/G)$
\item $\big(s\in\mathcal{L}(S_A/G)\big) \wedge\big( e^{|s|}_s \in \Sigma_o \big)\wedge\big( st_1 \in \lang(G)$ where $t_1\in \Sigma_{uo}^*(\Sigma_o\cup\{\epsilon\})\big) \wedge \big(\exists t_3 \in \hat{f}_A(P_o(s^{|s|-1})),\ \exists t_2 \in f_A(t_3,e^{|s|}_s)$ and $i_1\leq i_2\leq \hdots \leq i_{|t_1|} \in \mathbb{N}^{|t_2|} $ s.t. $\forall j \in \mathbb{N}^{|t_1|}: e^j_{t_1} \in S^d_A(t_3t_2^{i_j})\big)\wedge \big(P_o(t_1)\neq \epsilon \Rightarrow i_{|t_1|} = |t_2|\big)\ \Leftrightarrow st_1 \in \lang(S_A/G)$ 
\end{enumerate}
The above definition captures the intricate interaction between plant, supervisor and attacker.
Two important concepts are applied in this definition. 
First, the supervisor issues a control decision whenever it receives an observable event.
An observable event can be either a legitimate event or a fictitious event inserted by an attacker.
Second, the plant can execute any unobservable event enabled by the current control decision.
To demonstrate how to compute $\lang(S_A/G)$, we illustrate condition (2) in Figure \ref{fig:lang}.
\\
Assume that $uo_1,uo_2\in\Sigma_{uo}$, $ob\in\Sigma_o$ and $uo_1uo_2,$ $uo_1uo_2ob\in \lang(G)$.
Figure~\ref{fig:lang}(\subref{fig:langunobs}) describes how $t_1 = uo_1uo_2 \in \lang(S_A/G)$.
The string $uo_1uo_2$ belongs to $\lang(S_A/G)$ whenever there exists a string $t_2 \in f_A(\epsilon,\epsilon)$ and indices $i_1\leq i_2 \in \mathbb{N}$ such that $uo_1 \in S_A^d(t_2^{i_1})$ and $uo_2 \in S_A^d(t_2^{i_2})$.
Similarly, Fig.~\ref{fig:lang}(\subref{fig:langobs}) describes how $uo_1uo_2ob \in \lang(S_A/G)$.
If we use the same indices $i_1,i_2$ as before, then we just need to test if $ob \in S_A^d(t_2^{i_3})=S_A^d(t_2)$.
In the case of the last event being observable, we assume that the attacker has finished its entire modification $t_2$.
On the other hand, we assume that unobservable events can be executed based on control decisions of string prefixes of $t_2$.
Condition (3) applies the same mechanism of Condition (2), however it has nuances related to previous modifications made by the attacker.
\\
Similarly to the language $\mathcal{L}(S_A/G)$, we recursively define the set of reachable states of $G$ under the supervision of $S^d_A$ driven by the edited string $s$, for any $s \in (\Sigma_o\cup\Sigma_a^e)^*$, as follows:
\begin{align}
&RE^i_G(s,S_A^d) := \nonumber \\  
&\left\{ \begin{array}{l}\{x\in X | \exists u \in NX_{P_e^G(e^{i}_s)}\big(RE_G^{i-1}(s,S_A^d)\big),\\ \exists t \in (S_A^d(s^i)\cap\Sigma_{uo})^*: x = \delta(u,t)\}, \\ \hfill\text{if } e^{i}_s \in \Sigma_o\cup\Sigma_a^d \\
\{x\in X | \exists u \in \big(RE_G^{i-1}(s,S_A^d)\big),\\ \exists t \in (S_A^d(s^i)\cap\Sigma_{uo})^*: x = \delta(u,t)\}, \\ \hfill\text{if } e^{i}_s \in \Sigma_a^i
\end{array}\right.
\end{align}
for $ 1\leq i \leq |s|$ and:
\begin{align}
RE^0_G(s,S_A^d) := &\{
x\in X| \exists t \in (S^d_A(\epsilon)\cap\Sigma_{uo})^*: \nonumber \\& x=\delta(x_0,t)\}
\end{align}
We extended the definition under $S^d_A$ for edited strings since $S_A$ is nondeterministic. 
In the above definition, we want to define the reachable set of states \emph{after a particular edited string was selected}.
Note that, the definitions of $RE^i_G(s_A,S_A^d)$ and of $\lang(S_A/G)$ share many similarities since they are related.

The above-described dynamical interaction between the attacker and the supervisor is different than the one in \cite{Su:2018}, where the supervisor reacts to \emph{strings} of inserted events produced by the attacker.

\subsection{Attacker objectives}

\begin{figure}[thpb]
      \centering
      \includegraphics[width=.4\columnwidth]{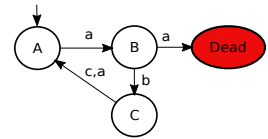}
      \caption{Supervisor $\tilde{R}$ for Example~\ref{ex1}}
      \label{fig:supervisor_R}
\end{figure}

In order to formally pose the problem, we must specify the objective of the attacker.
We assume that $G$ contains a set of \textit{critical} unsafe states defined as 
$X_{crit}\subset X$ such that $\forall x \in X_{crit}$, 
$x$ is never reached when $S_P$ controls $G$ and no attacker is present.
In general, not all states reached by strings of $G$ that are disabled by $S_P$ (when no attacker is present) are critically unsafe.
In practice, there will be certain states among those that correspond to physical damage to the system, such as ``overflow'' states or ``collision'' states, for instance.
Similar notions of  critical unsafe states have been used in other works, e.g., \cite{Paoli:2005,Paoli:2011}. 
Therefore, the objective of the attacker is to force the controlled behavior under attack $\mathcal{L}(S_A/G)$ to reach any state in $X_{crit}$.
\\
The second objective of the attacker is to remain stealthy, i.e., the attacker should feed the supervisor with \emph{normal} behavior.
By normal behavior, we mean that the supervisor should receive strings in the language $\lang(S_P/G)$.
We assume that if the supervisor receives a string that is not included in  $\lang(S_P/G)$, then an intrusion detection module detects the attacker.
\\
To capture the behavior that detects the attacker, we define an automaton that captures both the control decisions of the supervisor and the normal/abnormal behavior. 
We start by defining the automaton $H = obs(R||G)$, where $obs$ is the standard observer automaton of $G$ (cf. \cite{Lafortune:2008}) and $H = (X_H,\Sigma_o,\delta_H,x_{0,H})$, where $X_H\subseteq 2^{X_R\times X_G}$.
The automaton $H$ captures only the normal projected behavior of the controlled system. 
However, $H$ cannot be used as an supervisor since it only  could contain inadmissible control decisions and it does not have all decisions of $R$.
\\
Based on $H$, we define $\tilde{R} = (\tilde{Q},\Sigma,\tilde{\mu},\bar{q}_0)$, where $\tilde{Q}=X_H\cup\{$dead$\}$, and $\tilde{\mu}$ is defined to include all the transitions in $\delta_H$ plus the additional transitions: $(\forall q\in X_H)$~$(\forall e \in [(\Sigma_{uc}\cap\Sigma_o)\backslash\Gamma_H(q)])$ $\tilde{\mu}(q,e) =~$dead, $(\forall q\in X_H)$~$(\forall e \in (\Sigma_{uc}\cap\Sigma_{uo}))$ $\tilde{\mu}(q,e) = q$, $(\forall q\in X_H)$~$(\forall e \in (\Sigma_{c}\cap\Sigma_{uo})$ s.t. $\exists x\in q,\ e\in \Gamma_{R||G}(x))$ $\tilde{\mu}(q,e) = q$ and  $(\forall e \in \Sigma_{uc})\ \tilde{\mu}(\textrm{dead},e) = \textrm{dead}$.
In this manner, automaton $\tilde{R}$ embeds the same admissible control decisions as automaton $R$ and it differentiates normal/abnormal behavior. 
The state \emph{dead} in $\tilde{R}$ captures the abnormal behavior of the controlled system.
The attacker remains stealthy as long as $\tilde{R}$ does not reach the state \emph{dead}.
Figure~\ref{fig:supervisor_R} illustrates the supervisor $\tilde{R}$ computed for Example~\ref{ex1}.

\emph{Remark:} No new transition to the dead state with controllable events are included in the definition of $\tilde{R}$.
For simplicity, we assume that the supervisor does not enable controllable events unnecessarily.
In this manner, only uncontrollable events can reach the dead state.

\subsection{Problem formulation}

Finally, we are able to formally state the problem formulation of the synthesis of stealthy sensor deception attack problem.

\begin{prob}[Synt. of Stealthy Sensor Deception Attacks] \label{prob:main}
Given an attacker that has full knowledge of the models $G$ and $\tilde{R}$, and is capable of compromising events $\Sigma_a\subseteq\Sigma_o$, synthesize an attack function $f_A$ such that it generates a controlled language $\mathcal{L}(S_A/G)$ that satisfies:
\begin{enumerate}[align=parleft,labelsep=.4cm,leftmargin=*]
\item[1.] $\forall s \in P_o(\mathcal{L}(S_A/G))$, $\hat{f}_A(s)$ is defined \textbf{(Admissibility)};
\item[2.] $\forall s \in \mathcal{L}(S_A/G)$, $P_e^S\big(\hat{f}_A(P_{o}(s))\big) \subseteq  P_o(\mathcal{L}(S_P/G))$ \textbf{(Stealthiness)};
\item[3(a).] $\exists s \in \mathcal{L}(S_A/G)$ s.t.\ $\big(\forall t \in [P_{o}^{-1}(P_{o}(s))\cap\mathcal{L}(S_A/G)]\big)$  $\delta(x_0,t) \in X_{crit}$.
\end{enumerate}
In this case, we say that $f_A$ is a \textbf{strong attack}. 
We  additionally define the notion of a \textbf{weak attack} as follows:
\begin{enumerate}[align=parleft,labelsep=.4cm,leftmargin=*]
\item[3(b).] $\exists s \in \mathcal{L}(S_A/G)$ s.t.\ $\delta(x_0,s) \in X_{crit}$. 
\end{enumerate}
\end{prob}

The \textit{Admissibility} condition guarantees that $f_A$ is well defined for all projected strings in the modified controlled language $P_o(\mathcal{L}(S_A/G))$. 
The \textit{Stealthiness} condition guarantees that the attacker stays undetected by the supervisor, meaning that any string in $P_o(\mathcal{L}(S_A/G))$ should be modified to a string within the original controlled behavior. In this manner, $\tilde{R}$ never reaches state ``dead''.
Lastly, the reachability of critical states is stated in condition 3, where condition 3(a) is a strong version of the problem. 
In the strong case, the attacker is sure that the system has reached a critical state if string $s$ occurs in the system.
Condition 3(b) is a relaxed version, where the attacker might not be sure if a critical state was reached, although it could have been reached. 
Both variations of condition 3 guarantee the existence of at least one successful attack, namely, when string $s$ occurs in the new controlled behavior.

\emph{Remark:} Problem~\ref{prob:main} assumes that the attacker has full knowledge of the plant and the supervisor models. 
Although it might be difficult to achieve this assumption in practice, our paper studies the worst attack scenario case.
Moreover, this assumption is common practice within the cyber-security domain.

\subsection{Attack scenarios}

Problem \ref{prob:main} is defined for a general function $f_A$, which is defined as in Definition~\ref{def:attacker}. However,  there exists an interaction between the attacker, the system, and the supervisor that may limit the power of the attacker. 
For this reason, we propose three different types of attack functions. 
Each of them has different assumptions that are application-dependent.

The first scenario assumes that the attacker has ``time" to perform any \emph{unbounded} modification. 
In other words, the system does not execute any event until the attacker finishes its modification. 
Such an attack function is defined as an \emph{unbounded deterministic} attack function.

\begin{defn}
\label{def:unbounded}
An \textbf{unbounded} deterministic attack function is an attack function $f_A$ s.t. $\big(\forall s \in (\Sigma_o \cup \Sigma_a^e)^*\big)(\forall e\in\Sigma_o)[f_A(s,e)! \Rightarrow |f_A(s,e)| = 1]$. 
\end{defn}

The previous scenario considers a powerful attacker, and it might be unrealistic in many real applications. In this case, the second scenario limits the first one by considering only \emph{bounded deterministic} attack functions. In other words, we assume that the system does not react up to a bounded number of modifications made by the attacker. Bounded deterministic attack functions are defined next.

\begin{defn} 
\label{def:bounded}
Given $N_A \in \mathbb{N}^+$, a \textbf{bounded} deterministic attack function is an attack function $f_A$ s.t. $\big(\forall s \in (\Sigma_o \cup \Sigma_a^e)^*\big)(\forall e\in\Sigma_o)$ $[(f_A(s,e)! \Rightarrow |f_A(s,e)| = 1) \wedge (s_A \in f_A(s,e) \Rightarrow |s_A| \leq N_A )]$.  
\end{defn}

Finally, the last scenario is the least powerful attacker we consider. We assume that the system can interrupt the attacker's modification at any point. We call this function an \emph{interruptible} attack function. Formally, we define it as:

\begin{defn}
\label{def:interruptible}
An attack function $f_A$ is an \textbf{interruptible} attack function if $(\forall s \in (\Sigma_o\cup\Sigma_a^e)^*)(\forall e \in \Sigma_o)\ [s_A \in f_A(s,e) \Rightarrow \bar{s_A}\backslash \{\epsilon\} \subseteq f_A(s,e)]$.   
\end{defn}

We use a simple example to provide more intuition about the above-described scenarios.

\begin{exmp}
Let $\mathcal{L}(G) =\overline{ \{a\} \{b\}^*}$, 
where $\Sigma_a=\Sigma_o = \{ a, b\}$. %
Assume that $S_P$ never disables an event. 
We partially define three attack functions: $f^1_A, f^2_A$, and $f^3_A$.
\begin{align}
f^1_A(\epsilon,a) &= \{a\}\\ 
f^2_A(\epsilon,a) &= \{ab_i^n\},\text{ where }n\in \mathbb{N}\\
f^3_A(\epsilon,a) &= \{a,ab_i,ab_ib_i\}
\end{align}
$f_A^1$ is a bounded deterministic attack function with $N_A = 1$, $f^2_A$ is an unbounded deterministic attack function, and lastly, $f^3_A$ is an interruptible attack function.  
\end{exmp}

\section{Insertion-Deletion Attack Structure} \label{sec:IDA} 
\subsection{Definition}
An Insertion-Deletion Attack structure (IDA) is an extension of the notion of \textit{bipartite transition structure} presented in \cite{Yin:2016a}. 
An IDA captures the game between the environment and the supervisor considering the possibility that a subset of the sensor network channels may be compromised by a malicious attacker, whose moves will be constrained according to various rules. 
In this game we fix the supervisor's decisions as those defined in the given $\tilde{R}$. 
The environment's decisions are those of the attacker and the system. 
Therefore, in this game, environment states have both attacker's and system's decisions. 
In this section, we define the generic notion of an IDA.
In the next sections, we will construct specific instances of it, according to the permitted moves of the attacker. 

In order to build the game, we define  an information state as a \textit{pair} $IS \in 2^X\times \tilde{Q}$, and the set of all information states as $I = 2^{X}\times \tilde{Q}$. 
The first element in an $IS$ represents the \textit{correct state estimate} of the system, as seen by the attacker for the actual system outputs. 
The second element represents the supervisor's state, which is the current state of its realization based on the edited string of events that it receives. 
As defined, an $IS$ embeds the necessary information for either player to make a decision.
\begin{defn}\label{def:IDA}
An Insertion-Deletion Attack structure (IDA) $A$ w.r.t.\  $G$, $\Sigma_a$, and  $\tilde{R}$, is a 7-tuple
\begin{equation}
A = (Q_S,Q_E,h_{SE},h_{ES},\Sigma,\Sigma_a^e,y_0)
\end{equation}
where:
\begin{itemize}
\item $Q_S \subseteq I$ is the set of S-states, where S stands for Supervisor and where each S-state is of the form $y = (I_G(y),I_S(y))$,  where $I_G(y)$ and $I_S(y)$ denote the correct system state estimate and the supervisor's state, respectively;
\item $Q_E \subseteq I$ is the set of E-states, where E stands for Environment; each E-state is of the form $z = (I_G(z),I_S(z))$ defined in the same way as in the S-states case;
\item $h_{SE}:Q_S\times\Gamma\rightarrow Q_E$ is the partial transition function from S-states to E-states, defined only for $\gamma=\Gamma_{\tilde{R}}(I_S(y))$: 
\begin{gather}
\begin{split}
&h_{SE}(y,\gamma) := (UR_{\gamma}(I_G(y)),\ I_S(y))
\end{split}
\end{gather}
\item $h_{ES}:Q_E\times(\Sigma_o\cup\Sigma^e_a)\rightarrow Q_S$ is the partial transition function from E-states to S-states, satisfying the following constraints: for any $y\in Q_S$, $z\in Q_E$ and $e \in \Sigma_o\cup\Sigma_a^e$, if $h_{ES}(z,e)$ is defined, then $h_{ES}(z,e) := y$
where:
\end{itemize}
\vspace*{-0.2cm}
\begin{subnumcases}{y=}
 \big(NX_e(I_G(z)),\tilde{\mu}(I_S(z),e)\big), \textbf{ if } \nonumber \\ \hspace{1cm}e \in \Gamma_{\tilde{R}}\big(I_S(z)\big)\cap\Gamma_G\big(I_G(z)\big) \label{eq:hes1}
   \\ \nonumber\\
 {\big(I_G(z), \tilde{\mu}(I_S(z),P^S_e(e))\big),} \textbf{ if } e\in\Sigma_a^i\text{ and }\nonumber \\ \hspace{2cm} \mathcal{M}_e(e) \in \Gamma_{\tilde{R}}\big(I_S(z)\big)\label{eq:hes2}\\ 
  \nonumber\\
  {\big(NX_{P^G_e(e)}(I_G(z)),I_S(z)\big),} \nonumber  \textbf{ if } e\in\Sigma_a^d\text{ and } \\  \hspace{0.8cm}\mathcal{M}_e(e) \in \Gamma_{\tilde{R}}\big(I_S(z)\big)\cap\Gamma_G\big(I_G(z)\big)\label{eq:hes3}
\end{subnumcases}
\vspace*{-0.5cm}
\begin{itemize}
\item $\Sigma$ is the set of events of $G$;
\item $\Sigma^e_a$ is the set of editable events;
\item $y_0 \in Q_S$ is the initial S-state: 
$y_0 := ( \{ x_0\}, q_0)$.
\end{itemize}
\end{defn}

Since the purpose of an IDA is to capture the game between the supervisor and the environment, we use a bipartite structure to represent each entity. 
An S-state is an $IS$ containing the state estimate of the system $G$ and the supervisor's state; it is where the supervisor issues its control decision. 
An E-state is an $IS$ at which the environment (system or attacker) selects one among the observable events to occur. 

A transition from a S-state to an E-state represents the updated unobservable reach in $G$'s state estimate together with the current supervisor state. 
Note that $h_{SE}$ is only defined for $y$ and $\gamma$ such that $\gamma=\Gamma_{\tilde{R}}(I_S(y))$.
On the other hand,
a transition from an E-state to a S-state represents the ``observable reach'' immediately following the execution of the observable event by the environment. 
In this case, both the system's state estimate and the supervisor's state are updated. 
However, these updates depend on the type of event generated by the environment: 
(i) true system event unaltered by the attacker;
(ii) (fictitious) event insertion by the attacker; or 
(iii) deletion by the attacker of an event just executed by the system. 
Thus, the transition rules are split into three cases, described below.

The partial transition function $h_{ES}$ is characterized by three cases: Equations (\ref{eq:hes1}),(\ref{eq:hes2}), and (\ref{eq:hes3}). 
Equation (\ref{eq:hes1}) is related to \emph{system's} actions, while Equations (\ref{eq:hes2}) and (\ref{eq:hes3}) are related to \emph{attacker's} actions.
In the case of Equation (\ref{eq:hes1}), the system generates a feasible (enabled) event and the attacker lets the event reach the supervisor intact, either because it cannot compromise that event, or because it chooses not to make a move.
In the case of Equation (\ref{eq:hes2}), the attacker only inserts events consistent with the control decision of the current supervisor state.
In the case of Equation (\ref{eq:hes3}), it only deletes actual observable events generated by the system. 
Equation~(\ref{eq:hes3}) differs from Equation~(\ref{eq:hes1}) since it adds the condition that the event executed is compromised and that the attacker deleted it. 
\emph{Remark:} a given IDA will contain some attack moves (since it is a generic structure), all of which have to satisfy the constraints in the definition of $h_{ES}$.

In the remainder of this paper, we assume that all states included in an IDA are reachable from its initial state.

\begin{figure}[thpb]
	\centering
	\begin{subfigure}[t]{.45\columnwidth}
		\centering
		\includegraphics[width=.7\columnwidth]{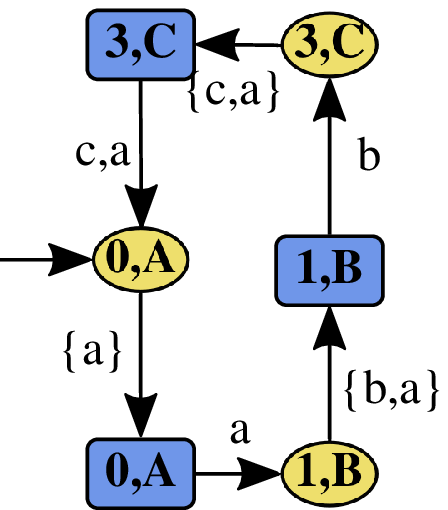}
		\caption{An IDA with no attacks}
		\label{fig:IDA1}
	\end{subfigure}%
	\begin{subfigure}[t]{.55\columnwidth}
		\centering
		\includegraphics[width=1\columnwidth]{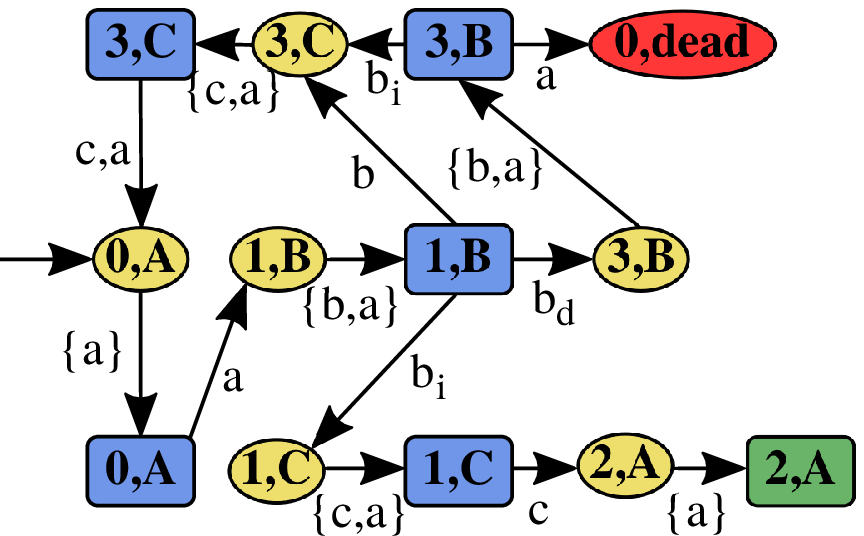}
		\caption{A second IDA example}
		\label{fig:IDA2}
	\end{subfigure}
	\caption{IDA Structures}
	\label{fig:IDA}
\end{figure}

\begin{exmp}\label{ex2}
Let us consider system $G$ and supervisor $\tilde{R}$ from Example \ref{ex1}. Considering the compromised event set  $\Sigma_a = \{b\}$, Fig.~\ref{fig:IDA} gives two IDA examples.
In the figure, oval states represent S-states and rectangular states represent E-states. 
Moreover, the red state indicates where the supervisor reaches the ``dead" state and the green state (2,A) represents the successful reaching of a critical state. 
\end{exmp}
%
%

Given two IDAs $A_1$ and $A_2$, we say that $A_1$ is a subsystem of $A_2$, denoted by $A_1 \sqsubseteq A_2$, if $Q_E^{A_1}\subseteq Q_E^{A_2}$,  $Q_S^{A_1}\subseteq Q_S^{A_2}$, and for any $y \in Q_S^{A_1}$, $z \in Q_E^{A_1}$, $\gamma \in \Gamma$, and $e\in\Sigma$, we have that:
\begin{enumerate}
\item $h_{SE}^{A_1}(y,\gamma) = z$ $\Rightarrow$ $h_{SE}^{A_2}(y,\gamma) = z$ and
\item $h_{ES}^{A_1}(z,e) = y$ $\Rightarrow$ $h_{ES}^{A_2}(z,e) = y$.
\end{enumerate}

Before, we discuss relevant properties of the IDA structure, we define a property about E-states.

(P1)~~An E-state $z \in Q_E$ is called a \textit{race-free state}, if the following condition holds:
\begin{equation*}
\label{eq:racecondition}
\begin{aligned}
&\forall e \in \Gamma_{\tilde{R}}(I_S(z))\cap\Sigma_o:(\exists x \in I_G(z):\delta(x,e)!)\Rightarrow \\ &[h_{ES}(z,e)! \vee h_{ES}(z,e_d)!]
\end{aligned}
\end{equation*}
Property (P1) ensures the non-existence of a race condition between the attacker and the system in a given E-state. 
Specifically, when (P1) holds in any E-state, the attacker has the option of either letting the event reach the supervisor intact or preventing the event from reaching the supervisor (or allowing both). 
It means that the attacker can wait for the system's response to the most recent control action and react accordingly. 
Note that if the event in question is a compromised event, then the attacker has the option of allowing both actions since we defined nondeterministic attack functions.
In the case of deterministic attack functions, we force the attacker to select one of the actions.
If (P1) does not hold at a particular E-state, then the attacker \emph{must insert an event}, and this insertion must take place before the system reacts to the most recent control action sent by the supervisor.
(In some sense, the attacker is ``racing'' with the system.)
An IDA that satisfies (P1) for all E-states is called a \textit{race-free} IDA.

\begin{defn}[Induced E-state] Given an IDA $A$, $IE(z,s)$ is defined to be the E-state induced by string $s\in (\Sigma_o\cup\Sigma_a^e)^*$, when starting in the E-state $z$. $IE(z,s)$ is computed recursively as:
\begin{equation*}
\begin{aligned}
IE(z,\epsilon) &\coloneqq z \\
IE(z,se) &\coloneqq \left\{
\begin{array}{ll}
	h_{SE}\big(y,\Gamma_{\tilde{R}}\big(I_S(y)\big)\big)& \textrm{if~} h_{ES}(IE(z,s),e)! \\ 
	\text{undefined} & \textrm{otherwise}
\end{array}
\right.
\end{aligned}
\end{equation*}
where $y=h_{ES}(IE(z,s),e)$
\end{defn}
We also define $IE(s)\coloneqq IE(z_0,s)$, where $z_0 = h_{SE}(y_0,\Gamma_{\tilde{R}}(I_S(y_0)))$.

We conclude this section by defining the notion of  \emph{embedded} attack function in an IDA and presenting two lemmas about reachability in IDAs. 
The first lemma shows that after observing an edited string $s_A$, the attacker correctly keeps track of the supervisor state $\tilde{R}$. 
Then we show that an IDA correctly estimates the set of possible states of the system after the occurrence of edited string $s_A$.
\begin{defn}[Embedded $\mathbf{f_A}$]\label{def:fa_inclusion} An attack function $f_A$ is said to be embedded in an IDA $A$ if 
$\big(\forall s \in P_o(\mathcal{L}(S_A/G))\big) \allowbreak \big(\forall t \in \hat{f}_A(s)\big) \big( \forall i \in \mathbb{N}^{|t|-1}\big)$, then $IE(t^i)$ is defined. 
%
\end{defn}
Intuitively, an attack function is embedded in an IDA if for every modified string that is consistent with the behavior of $G$, we can find a path in the IDA for that string.
Note that we limit this constraint on $f_A$ to strings in $P_o(\mathcal{L}(S_A/G))$, since these are the ones consistent with G under a given $f_A$.
We are not interested in the definition of any $f_A$  for strings outside of the controlled behavior.
\begin{lem}\label{lemma:supervisor_estimate}
Given a system $G$, supervisor $\tilde{R}$, and an IDA structure $A$ with an embedded attack function $f_A$,  for any string $s\in\mathcal{L}(S_A/G)$ and any string $s_A \in \hat{f}_A(P_o(s))$, we have
\begin{equation}
I_S\big(IE(s_A)\big) = \tilde{\mu}\big( q_0,P_e^S(s_A)\big)
\end{equation}
\begin{equation}
I_G\big(IE(s_A)\big) = RE^{\lvert s_A\rvert}_G\big(s_A,S_A^d\big)
\end{equation}
%
%
%
\end{lem}

\section{All Insertion-Deletion Attack Structure}\label{sec:AIDA}

In this section, we define the \emph{All} Insertion-Deletion Attack Structure, a specific type of IDA abbreviated as AIDA hereafter. 
Given $S_P$ and $\Sigma_a$, the AIDA embeds \emph{all} insertion-deletion actions the attacker is able to execute. 
We discuss its construction and properties.

\subsection{Definition}

As consequence of Lemma \ref{lemma:supervisor_estimate}, if we construct an IDA structure based on $S_P$ and $\Sigma_a$ that is ``as large as possible", then it will include all valid insertion and deletion actions for the attacker. 
We formally define such a structure as the All Insertion-Deletion Attack structure.

\begin{defn}[$\mathsf{AIDA}(G,S_P,\Sigma_a)$] \label{def:AIDA} Given a system $G$, a supervisor $S_P$, and a set of compromised events $\Sigma_a$, the \textbf{All Insertion-Deletion Attack structure (AIDA)}, denoted by $\mathsf{AIDA}(G,S_P,\Sigma_a) = (Q_S,Q_E,h_{SE},h_{ES},\Sigma,\Sigma_a^e,y_0)$, is defined as the \textbf{largest} IDA w.r.t to $G$, $S_P$ and $\Sigma_a$ s.t.
\begin{enumerate}
\item For any $y \in Q_S$, we have $|\Gamma_{AIDA}(y)| = 0 \Leftrightarrow I_S(y) =~$dead
\item For any $z \in Q_E$, we have
\begin{enumerate}
\item
$\forall e\in \Gamma_{\tilde{R}}(I_S(z))\cap\Gamma_G(I_G(z))\cap\Sigma_o: (h_{ES}(z,e)! \vee  h_{ES}(z,e_d)!)$ or
\item $I_G(z) \subseteq X_{crit} \Rightarrow |\Gamma_{AIDA}(z)| = 0$
\end{enumerate}
\end{enumerate}
\end{defn}

Condition 2(a) alone satisfies the non-existence of a race condition in the AIDA.
Conditions 1 guarantees that the AIDA stops its search once the attack is detected.
Similarly, Condition 2(b) stops the search given that the attacker knows it has reached its goal.

By ``largest" structure, we mean that for any IDA $A$ satisfying the above conditions: $A\sqsubseteq \mathsf{AIDA}(G,S_P,\Sigma_a)$.
This notion of ``largest" IDA is well defined. 
If $A_1$ and $A_2$ are two IDA structures satisfying the above conditions, then their union still satisfies the conditions, where the union $A_1\cup A_2$ is defined as: $Q^{A_1\cup A_2}_E=Q_E^{A_1}\cup Q_E^{A_2}$, $Q^{A_1\cup A_2}_S = Q_S^{A_1}\cup Q_S^{A_2}$, and for any $y\in Q^{A_1\cup A_2}_E$, $z\in Q^{A_1\cup A_2}_S$, $\gamma \in \Gamma$ and $e\in\Sigma\cup\Sigma_a^e$, we have that $h_{SE}^{A_1\cup A_2}(y,\gamma) = z \Leftrightarrow \exists i\in\{1,2\}: h_{SE}^{A_i}(y,\gamma) = z$ and $h_{ES}^{A_1\cup A_2}(z,e) = y \Leftrightarrow \exists i\in\{1,2\}: h_{ES}^{A_i}(z,e) = y$.

\subsection{Construction}
The construction of the AIDA follows directly from its definition. 
It enumerates all possibles transitions for each state by a breadth-first search. 
Each S-state has at most one control decision, which is related to the supervisor state $\tilde{R}$. 
On the other hand, each E-state enumerates both system's and attacker's  actions, according to its system and supervisor estimate, respectively. 
In practice, we do not need to search the entire state space; we stop the branch search when an S-state reaches a state with the supervisor at the ``dead" state or when the system estimate of an E-state is a subset of $X_{crit}$.

The procedure mentioned above is described in Algorithm \ref{algo:construction} (Construct-AIDA). 
It has the following parameters: $AIDA$ is the graph structure of the AIDA we want to construct; $AIDA.E$ and $AIDA.S$ are the E- and the S-state sets of the structure, respectively; $AIDA.h$ is its transition function; $Q$ is a queue. 
We begin the procedure by initializing $AIDA.S$ with a single element $y_0 = (\{x_0\},q_0)$. 
The breath-first search is then performed by the procedure DoBFS. 
The transitions between S-states and E-states are dealt within lines 7 to 11. 
The transitions between E-states and S-states are defined in lines 12 to 25, where each attack possibility is analyzed. 
These transitions are defined exactly as in Definition \ref{def:IDA}, equations (\ref{eq:hes1}), (\ref{eq:hes2}) and (\ref{eq:hes3}). 
(For the sake of readability, we employ the usual triple notation (origin, event, destination) for the transition function.)
The procedure converges when all uncovered states (states in the queue) are covered, meaning we have traversed the whole reachable space of E- and S-states. 
Note that lines $29$ and $33$ impose the stop conditions of  Definition \ref{def:AIDA}.

\begin{algorithm}
\caption{Construct-AIDA}
\label{algo:construction}
\begin{algorithmic}[1]
\Require $G$, $\tilde{R}$ and $\Sigma_a$
\Ensure $AIDA$
	\State $AIDA$ $\leftarrow$ DoBFS$\big(G$, $\tilde{R}$, $(\{x_0\},q_0)\big)$
	\Procedure{DoBFS}{$G,R,y$}
		\State $AIDA.S\leftarrow \{y\}$, $AIDA.E\leftarrow\emptyset$, $AIDA.h\leftarrow\emptyset$
		\State Queue $Q \leftarrow \{y\}$
		\While{$Q$ is not empty}
			\State $c \leftarrow Q.dequeue(\ )$
			\If{$c \in AIDA.S$}
				\State $\gamma \leftarrow \Gamma_{\tilde{R}}(I_S(c))$
				\State $z\leftarrow\big(UR_{\gamma}\big(I_G(c)\big),I_S(c)\big)$
				\State $AIDA.h \leftarrow AIDA.h\cup\{(c,\gamma,z)\}$
				\State Add-State-to-AIDA$(z,AIDA,Q)$
				
			\ElsIf{$c \in AIDA.E$}
				\ForAll{$e\in\Sigma_o\cap\Gamma_{\tilde{R}}\big(I_S(c)\big)$}
					\If{$e \in \Gamma_G\big(I_G(c)\big)$}
						\State $y\leftarrow \Big(NX_e\big(I_G(c)\big),\tilde{\mu} \big(I_S(c),e\big)\Big)$
						\State $AIDA.h \leftarrow AIDA.h\cup \{(c,e,y)\}$
						\State Add-State-to-AIDA$(y,AIDA,Q)$
					\EndIf
					\If{$e\in\Gamma_G\big(I_G(c)\big)\wedge e\in\Sigma_a$}
						\State $y\leftarrow \big(NX_e\big(I_G(c)\big), I_S(c)\big)$
						\State $AIDA.h \leftarrow AIDA.h\cup \{(c,e_d,y)\}$
						\State Add-State-to-AIDA$(y,AIDA,Q)$
					\EndIf
					\If{$e\in\Sigma_a$}
						\State $y\leftarrow \big(I_G(c), \tilde{\mu} \big(I_S(c),e\big)\big)$
						\State $AIDA.h \leftarrow AIDA.h\cup \{(c,e_i,y)\}$
						\State Add-State-to-AIDA$(y,AIDA,Q)$
					\EndIf
				\EndFor
			\EndIf
		\EndWhile
	\EndProcedure
	\Procedure{Add-State-to-AIDA}{$c,AIDA,Q$}
		\If{$c \notin AIDA.E$ $\wedge\ c$ is an E-state}
					\State $AIDA.E \leftarrow AIDA.E\cup \{c\}$
					\If{$I_G(c)\nsubseteq X_{crit}$}		
						\State $Q.enqueue(c)$
					\EndIf	
		\ElsIf{$c\notin AIDA.S \wedge\ c$ is a S-state}
			\State $AIDA.S\leftarrow AIDA.S\cup \{c\}$	
			\If{$I_S(c)\neq$ dead}		
				\State $Q.enqueue(c)$
			\EndIf				
		\EndIf		
	\EndProcedure
\end{algorithmic}
\end{algorithm}

\begin{thm}\label{theo:AIDA}
 Algorithm Construct-AIDA correctly constructs the AIDA.
%
\end{thm}

\begin{exmp}\label{ex:AIDA}
We return to system $G$ and supervisor $\tilde{R}$ in Fig.~\ref{fig:ex1}(\subref{fig:IDA2}) with $\Sigma_a=\{b\}$. The IDA shown in Fig.~\ref{fig:IDA}(\subref{fig:IDA2}) is its AIDA. Note that there exists an $S$-state where $\tilde{R}$ reaches state \emph{dead}. Moreover, there exists an $E$-state where $G$ reaches a critical state, $(2,A)$.
\end{exmp}
\emph{Remark:} The AIDA has at most $2^{|X|+1}|\tilde{Q}|$ states since $Q_1,Q_2\subseteq I$.
If $\Sigma_{uo} = \emptyset$, then it has at most $2|X||\tilde{Q}|$ states.
\section{Synthesis of Stealthy IDA: Interruptible Case} \label{sec:synthesis-inter} 
The AIDA embeds all attack functions, including non-stealthy strategies, i.e., those that lead to state \emph{dead} of the supervisor. 
In this section, we show how to synthesize \emph{interruptible} and \emph{stealthy} attack functions that solve Problem \ref{prob:main}. First, we present a pruning process that removes non-stealthy interruptible strategies from the AIDA. 
The resulting pruned IDA is then used in the synthesis algorithm. 
Recall that three attack types were presented in Section~\ref{sec:deception} (Definitions~\ref{def:unbounded}, \ref{def:bounded}, and \ref{def:interruptible}).
We focus our attention on the interruptible case in this section; next, in Section \ref{sec:othersynthe}, we present our results for the two remaining cases.
\subsection{Pruning Process}

The AIDA could reach state \emph{dead} of supervisor $\tilde{R}$.
Each time this occurs, it means that the last step of the attack is no longer stealthy.
Hence, we must prune the AIDA in order to embed only stealthy attacks.
We pose this pruning process as a \emph{meta-supervisory-control} problem, where the ``plant'' is the entire AIDA, the specification for that plant is to prevent reaching state \emph{dead} of the supervisor, and the \emph{controllable} events are the actions of the attacker.
We assume that the reader is familiar with the standard ``Basic Supervisory Control Problem" of \cite{Ramadge:1987}; we adopt the presentation of that problem as BSCP in \cite{Lafortune:2008}.
First, we show necessary modifications to the standard BSCP algorithm in order to construct the Stealthy-AIDA.

\begin{algorithm}
\caption{Modified BSCP for Interruptible Attacker}
\label{algo:BSCP}
\begin{algorithmic}[1]
	\Require $A = (Q_E\cup Q_S,E,f_{se}\oplus f_{es},a_0= y_0)$, where $E \subseteq (\Sigma_o\cup\Sigma_a^e\cup\Gamma)$ and $A_{trim}= (A^t,E,f^t,a_0^t)$, where $A^t \subseteq Q_E\cup Q_S$
	\State\textbf{Step 1} Set $H_0 = (A_0,E,g_0,a_0) =  A_{trim}$, and $i=0$
	\State\textbf{Step 2} Calculate
	\State \textbf{Step 2.1} $A_i'= \{ a \in A_i| \Gamma_{A} (a)\cap E_{uc}\subseteq\Gamma_{H_i}(a)\}$ 
	\State\textbf{Step 2.2} $A_i^*=\{a \in A_i'| e \in \Gamma_A (a) \Rightarrow (e\in \Gamma_{H_i}(a) \vee e_d\in\Gamma_{H_i} (a)\}$

	$g_i'= g_i|A_i^*$  [transition function update]
	\State\textbf{Step 2.3} $H_{i+1} = Trim(A_i^*,E,g_i',a_0)$
	\State\textbf{Step 3} If $H_{i+1} = H_i$, Stop; otherwise $i\leftarrow i+1$, back to Step $2$
\end{algorithmic}
\end{algorithm}
%
The difference between the original BSCP algorithm (see, e.g., \cite{Lafortune:2008}) and its modified version in Algorithm \ref{algo:BSCP} is the addition of Step 2.2.
Since the BSCP algorithm has quadratic worst-time complexity in the number of states of the automaton $A_{trim}$, it follows that Algorithm~\ref{algo:BSCP} also has quadratic worst-time complexity.
In order to ensure the desired interruptibility condition of an attack function, we need that each state that it visits in the AIDA be race free.
%
Therefore, at Step 2.2 we enforce the race-free condition at every E-state of the IDA.
In order to enforce such condition, the algorithm deletes E-states where both the ``let through'' transition \emph{and} the ``erasure'' transition are absent for a feasible system event. 
Therefore, the resulting IDA from Algorithm \ref{algo:BSCP} is a race-free IDA.

To compute the stealthy AIDA structure, we define as system the AIDA constructed according to Algorithm \ref{algo:construction}. 
Moreover, any event $e \in \Sigma_a\cup\Sigma_a^e$ is treated as \emph{controllable} while any control decision $\gamma \in \Gamma$ and any event $e\in\Sigma_o\backslash\Sigma_a$ is treated as \emph{uncontrollable}.
The specification language, realized by $A_{trim}$, is obtained by deleting the states where the supervisor reaches the dead state, i.e., by deleting in the AIDA all states of the form $y = (S,$ dead$)$ for any $S\subseteq X$.


We formalize the pruning process for obtaining all stealthy insertion-deletion attacks as follows.

\begin{defn}\label{def:ISDA} Given the AIDA constructed according to Algorithm \ref{algo:construction}, define the system automaton $A_G = AIDA$ with $\Sigma^A_c = \Sigma_a\cup\Sigma_a^e$ as the set of controllable events and $\Sigma^A_{uc} = (\Sigma_o\backslash\Sigma_a)\cup\Gamma$ as the set of uncontrollable events. The specification automaton is defined by $A_{trim}$, which is obtained by trimming from $A_G$ all its states of the form $(S,$ dead$)$, for any $S\subseteq X$. The  \emph{Stealthy AIDA structure}, called the ISDA (Interruptible Stealthy Deceptive Attack), is defined to be the automaton obtained after running  Algorithm \ref{algo:BSCP} 
on $A_{trim}$ with respect to $A_G$ and $\Sigma^A_c$.
\end{defn}

\begin{figure}[thpb]
	\centering
	\begin{subfigure}[t]{.5\columnwidth}
		\centering
		\includegraphics[width=.8\columnwidth]{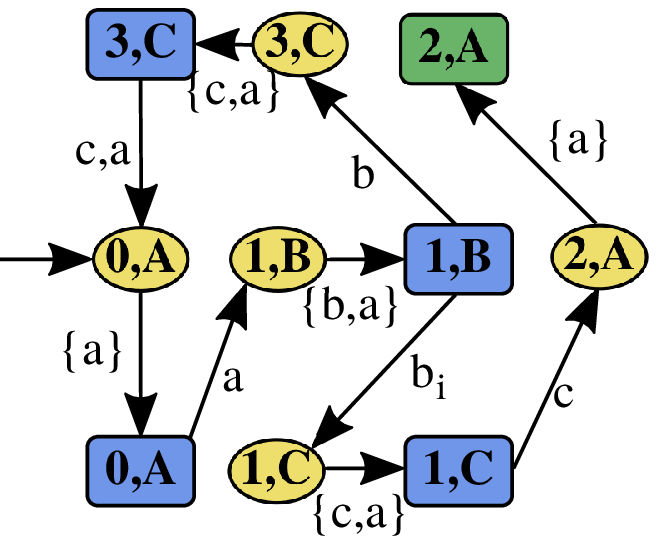}
		\caption{ISDA}
		\label{fig:ISDA}
	\end{subfigure}%
	\begin{subfigure}[t]{.5\columnwidth}
		\centering
		\includegraphics[width=1\columnwidth]{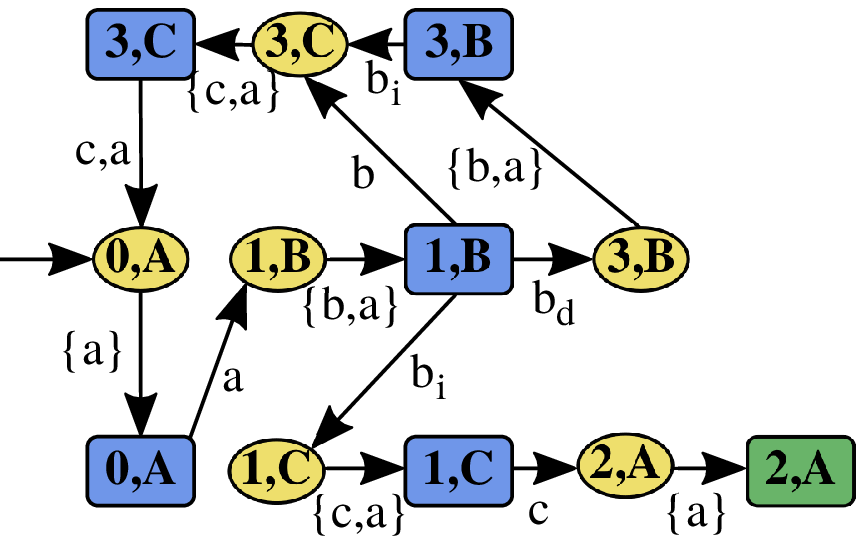}
		\caption{USDA}
		\label{fig:USDA}
	\end{subfigure}
	\caption{Stealthy Deceptive Structures}
	\label{fig:StIDA}
\end{figure}

\begin{exmp}
We return to the AIDA in Figure \ref{fig:IDA}, where we would like to obtain the ISDA as in Definition \ref{def:ISDA}. Figure \ref{fig:StIDA}(\subref{fig:ISDA}) shows the resulting ISDA, where the attacker cannot play event $b_d$ at $E$-state $(1,B)$. For example, if the attacker takes such decision, then the uncontrollable event $a$ could be executed, revealing the attack to the supervisor. 
\end{exmp}

\begin{lem}\label{lem:f_AinISDA}
If an interruptible attack function $f_A$ satisfies the admissibility and the stealthy conditions from Problem \ref{prob:main}, then it is embedded in the ISDA.
\end{lem}
\begin{lem}
\label{lem:f_AsyntesisISDA}
If an interruptible attack function $f_A$ is constructed from the ISDA, then the stealthy condition of Problem \ref{prob:main} is satisfied. 
\end{lem}
Constructing an $f_A$ from the ISDA means selecting decisions at E-states and properly defining $f_A$ from the selected decisions. 
\textit{Remark}: Lemma \ref{lem:f_AsyntesisISDA} does not guarantee the admissibility condition. An inadmissible interruptible $f_A$ could be synthesized from the ISDA. However an admissible interruptible $f_A$ can always be synthesized from the ISDA. 
Since the ISDA is race-free, an inadmissible $f_A$ synthesized from the ISDA can always be extended to make it admissible; see Example \ref{ex:extract} below. 
\begin{thm} \label{thm:all} The ISDA embeds all possible interruptible stealthy insertion-deletion attack strategies with respect to $\Sigma_a$, $\tilde{R}$ and $G$.
\end{thm}
%
\subsection{Synthesis of Stealthy Interruptible Functions}
Based on the ISDA, we can synthesize interruptible attack functions that satisfy the admissibility and the stealthy conditions. In order to fully satisfy Problem \ref{prob:main}, we need to address its last condition about the existence of strong or weak attacks. 
\begin{thm}\label{thm:main} 
Given the ISDA, there exists an interruptible $f_A$  that \emph{strongly} satisfies Problem \ref{prob:main} if and only if there exists an E-state $z$ in the ISDA s.t.\ $I_G(z) \subseteq X_{crit}$. 
On the other hand, it \emph{weakly} satisfies Problem \ref{prob:main} if and only if there exists an E-state $z$ in the ISDA s.t.\ $\exists x \in I_G(z)$, $x \in X_{crit}$.
\end{thm}
Theorem \ref{thm:main} gives necessary and sufficient conditions for synthesis of interruptible attack functions. The following algorithm (Algorithm \ref{algo:synthesis}) synthesizes a simple interruptible attack function that satisfies Problem \ref{prob:main}.
Specifically, Algorithm \ref{algo:synthesis} encodes an interruptible attack function in an automaton $F$. 
The encoded function simply includes \emph{one} attempt to reach a given critical state. 
First, it computes the shortest path from the initial state to a critical state via the function \textit{Shortest-Path$(ISDA,z\in Q_E^{ISDA})$} such that $I_G(z)\subseteq X_{crit}$ (strong attack) or $I_G(z)\cap X_{crit}\neq \emptyset$ (weak attack).
The second step is to expand the function, based on the shortest path, in order to satisfy the admissibility condition.
\begin{exmp}\label{ex:extract}
Let us extract an interruptible $f_A$ from the ISDA shown in Fig. \ref{fig:StIDA}(\subref{fig:ISDA}). 
The shortest path starting from the initial state $(0,A)$ to E-state $(2,A)$ goes through E-states $(0,A), (1,B)$ and $(1,C)$. 
Based on this path $f_A$ is defined, for example $f_A(\epsilon,a) = \{a, ab_i\}$. However, defining $f_A$ solely based on this path makes it inadmissible. 
The string $ab$ is defined in $\mathcal{L}(S_A/G)$ but $f_A(a,b)$ is undefined. 
For this reason, we have to expand the function $f_A$ after defining it for the shortest path. 
In this example, we first expand it for $f_A(a,b)  = \{b\}$, and later on for $f_A(ab,a) = \{a\}$ and $f_A(ab,c) = \{c\}$. Such extensions suffice to make $f_A$ admissible. 
\end{exmp}
\begin{algorithm}
\caption{Synthesis-$f_A$}
\label{algo:synthesis}
\begin{algorithmic}[1]
\Require $ISDA$, $z\in Q_E^{ISDA}$ s.t. $I_G(z)\subseteq X_{crit}$ or $I_G(z)\cap X_{crit}\neq \emptyset$
\Ensure Encoded $f_A$ in automaton $F$
\State $path \leftarrow$ Shortest-Path$(ISDA,z)$
\State $F\leftarrow$Expand-Path$(path,ISDA)$
\Procedure{Expand-Path}{$path,ISDA$}
\State Queue $Q\leftarrow \emptyset$; $F\leftarrow \emptyset$
\State $F.X \leftarrow \{q_0\}$; $F.x_0 = q_0$
\State $F.\delta \leftarrow \emptyset$; $F.\Sigma = \Sigma \cup \Sigma_a^e$
\ForAll{$(q,e,f) \in path \wedge q$ is an $E$-State}
	\State Add-to-F$(q,e,f,Q,A)$
 \EndFor
\While{Q is not empty}
	\State $q \leftarrow Q.dequeue()$
	\ForAll{$(q,e,f) \in ISDA.h$ $\wedge\ (q,e,f) \notin A.h$}
		\If{$e\in\Sigma_o$ $\wedge\ \nexists f^*$ s.t. $(q,e_d,f^*)\in A.h$}
			\State Add-to-F$(q,e,f,Q,A)$
		\ElsIf{$e\in\Sigma_a^d\wedge (q,\mathcal{M}(e),f^*)\notin ISDA.h$ }
			\State Add-to-F$(q,e,f,Q,A)$
		\EndIf
	\EndFor
\EndWhile
\EndProcedure
\Procedure{Add-to-F}{$q,e,f,Q,A$}
	\State $F.\delta \leftarrow F.\delta \cup \{(q,e,f)\}$
	\If{$f\notin Q$}
		\State $Q.enqueue(f)$
		\State $F.X \leftarrow F.X \cup \{f\}$
	\EndIf
\EndProcedure
\end{algorithmic}
\end{algorithm}
\textit{Remark}: We presented a simple synthesis algorithm.
Clearly, one could choose another strategy to extract an interruptible function from the ISDA. 
The important point here is that the ISDA provides a representation of the desired ``solution space'' for the synthesis problem.
We are not focused on the synthesis of minimally invasive attack strategies as studied in \cite{Su:2018}, where minimally invasive means an attack strategy with the least number of edits to reach a critical state.
\section{Other Attack Scenarios} \label{sec:othersynthe} 
In this section we present modifications to Algorithm \ref{algo:BSCP} in order to compute similar structures as the ISDA for the remaining two attack functions investigated in this paper. 
We omit the respective synthesis algorithms since they are similar to Algorithm \ref{algo:synthesis}. 
Although the ``expand path'' function of Algorithm \ref{algo:synthesis} changes for each attack function, such changes follows the assumption of each attack function.
\subsection{Deterministic Unbounded Attacks}
Different from the interruptible attack, in the deterministic unbounded (det-unb) attack case, we do not need to consider that the system may interrupt during an attack insertion. 
The attacker can insert events, possibly an arbitrarily long string in fact, before the system reacts.
As consequence, the pruned IDA for the det-unb attack is not necessarily race free. 

Algorithm \ref{algo:BSCP} prunes the AIDA enforcing it to be race free (Step $2.2$); however this condition needs to be relaxed for the det-unb case, resulting in  Algorithm \ref{algo:Det-Unb-BSCP}.
Step $2.1$ is also modified since we also need to relax the controllability condition.
Specifically, Step $2.1$ relaxes the controllability condition because the attacker ``races" with the system at states that violate this condition.\footnote{It is interesting to mention the similarity of ``racing" in our work with the work on supervisory control with forced events \cite{Golaszewski:1987}.} 
Algorithm \ref{algo:Det-Unb-BSCP} flags states that violate the controllability condition to later analyze if they need to be pruned or not. 
Note that once a state is flagged, it remains flagged throughout the algorithm. 
Step $2.2$ is divided into three steps. 
First, deadlocks created by the pruning process are deleted. 
Second, we flag all states violating the race-free condition. 
Then, the transition function is updated based on the flagged states. 
This update is such that only insertions transitions are possible from flagged states, since the attack will not wait for a reaction of the system. 
\begin{algorithm}
\caption{Det-Unb Modification}
\label{algo:Det-Unb-BSCP}
\begin{algorithmic}[1]
	\State \textbf{Step 2.1} Flag all $ a \in A_i$ s.t. $\Gamma_A (a)\cap E_{uc} \not\subseteq\Gamma_{H_i} (a)$  
	\State\textbf{Step 2.2}
	\State\textbf{	Step 2.2.1} $A_i^*=\{a \in A_i|\ \Gamma_{H_i}\big(a\big)=\emptyset \Rightarrow \Gamma_{A}(a)= \emptyset\}$
	\State\textbf{	Step 2.2.2} Flag all $a \in A_i^*$ s.t. $ \big(e \in \Sigma_o \wedge e \in \Gamma_A (a)\big) \Rightarrow \big( \{e, e_d\} \cap \Gamma_{H_i}(a) = \emptyset\big)$
	\State\textbf{	Step 2.2.3} For $a \in A_i^*$ and $e \in E$ 
	\vspace*{-0.5cm}	
	\[
	g_i'(a,e) = \left\lbrace 
	\begin{array}{ll}
		g_i(a,e) & \text{if } e \in (\Sigma_a^i \cup \Gamma) \wedge\\
		& g_i(a,e)! \\
		g_i(a,e) & \text{if } e \in (\Sigma_a^d \cup \Sigma_o) \wedge g_i(a,e)! \wedge\\
		& a \text{ not flagged}\\
		\text{undefined} & \text{otherwise}
	\end{array}		
	\right.\]
	\vspace*{-0.5cm}	
\end{algorithmic}
\end{algorithm}
We can adapt Definition \ref{def:ISDA} to prune the AIDA for the case of det-unb attacks by considering the modification of Step $2.1$ and Step $2.2$, as presented in Algorithm \ref{algo:Det-Unb-BSCP}. 
We name the resulting stealthy IDA as the USDA, for \textit{Unbounded Stealthy Deceptive Attack} structure.  Versions of Lemmas \ref{lem:f_AinISDA},~\ref{lem:f_AsyntesisISDA}, and Theorem~\ref{thm:all} are created for this specific attacker.
\begin{lem}\label{lem:f_AinUSDA}
A deterministic unbounded attack function $f_A$ is embedded in the USDA if it satisfies conditions (1) and (2) from Problem \ref{prob:main}.
\end{lem}
\begin{lem}\label{lem:f_AsyntesisUSDA}
If a det-unb attack function $f_A$ is synthesized from USDA, then the stealthy condition of Problem \ref{prob:main} is satisfied.
\end{lem}
\begin{thm} \label{thm:all_USDA} The USDA embeds all possible det-unb stealthy insertion-deletion attack strategies with respect to $\Sigma_a$, $R$ and $G$.
\end{thm}
\begin{exmp}\label{ex:USDA}
As we did for the ISDA, we also show the USDA structure for the AIDA in Figure \ref{fig:IDA}. Figure \ref{fig:StIDA}(\subref{fig:USDA}) shows the USDA, where the only deleted state is $(0,dead)$. Note that decision $b_d$ was deleted in the ISDA at state $(1,B)$, however it is maintained in the USDA. It is only possible since we know that decision $b_i$ can be played before the execution of event $a$ at state $(3,B)$. 
\end{exmp}
\subsection{Deterministic Bounded Attacks}

\begin{figure}[thpb]
	\centering
	\includegraphics[width=.5\linewidth]{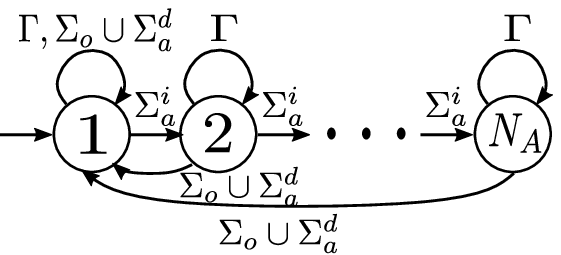}
	\caption{$G_{bound}$}
	\label{fig:bound}
\end{figure}

The AIDA structure is general enough for the interruptible and the unbounded attack scenarios; however, as constructed, it does not capture the bound in the case of deterministic bounded attacks (or det-bounded case).
We now present a simple mechanism in order to computed a bounded version of the AIDA, that we term BAIDA.
(The $||$ operation is the standard parallel composition of automata.)
\begin{defn}\label{def:BAIDA} Given the AIDA constructed by Algorithm \ref{algo:construction} and the automaton shown in Figure \ref{fig:bound}, the BAIDA is defined as $BAIDA = AIDA || G_{bound}$. 
(For the purpose of $||$, the AIDA is treated as an automaton.)
\end{defn}

The det-bounded attack case has similar conditions as the previously-discussed det-unb case, however, it can only perform bounded modifications. 
We need to take into account how many modifications the attacker has already performed at a given E-state. Therefore, synthesis algorithms for det-bounded attacks must use the BAIDA, as in Definition \ref{def:BAIDA}.

The BAIDA, as the AIDA previously, has to be pruned. 
We only show the modification of Step $2$ for Algorithm \ref{algo:BSCP}, resulting in  Algorithm \ref{algo:Det-Bounded-BSCP}. 
\begin{algorithm}
\caption{Det-Bounded Modification}
\label{algo:Det-Bounded-BSCP}
\begin{algorithmic}[1]
	\State \textbf{Step 2.1}
	\State \textbf{	Step 2.1.1} Flag all $ (a,n) \in A_i$ s.t. $\Gamma_A (a)\cap E_{uc} \not\subseteq\Gamma_{H_i} ((a,n))\wedge n<N_A$ 
	\State \textbf{	Step 2.1.2} $A_i'= \{ (a,n) \in A_i|\ n = N_A \Rightarrow \Gamma_A (a)\cap E_{uc}\subseteq\Gamma_{H_i} ((a,n)) \}$ 
	\State\textbf{Step 2.2}
	\State\textbf{	Step 2.2.1} $A_i''=\{(a,n) \in A_i'|\ \Gamma_{H_i}\big((a,n)\big)=\emptyset \Rightarrow \Gamma_{A}(a)= \emptyset\}$
	\State\textbf{	Step 2.2.2} $A_i^*=\{(a,n) \in A_i''|\ n = N_a \wedge \ e \in \Sigma_a \wedge e \in \Gamma_A (a) \Rightarrow \big(e\in \Gamma_{H_i}((a,n)) \vee e_d\in\Gamma_{H_i} ((a,n))\big)\}$
	\State\textbf{	Step 2.2.3} Flag all $(a,n) \in A_i^*$ s.t. $n<N_a$, $(e \in \Sigma_o \wedge e \in \Gamma_A (a)) \Rightarrow \big( \{e, e_d\}\cap\Gamma_{H_i}((a,n))=\emptyset\big)$
	\State\textbf{	Step 2.2.4} For $(a,n) \in A_i^*$ and $e \in E$
	\vspace*{-0.5cm}
	\[
	\hspace*{-1.5cm}g_i'((a,n),e) = \left\lbrace 
	\begin{array}{ll}
		g_i((a,n),e) & \text{if } e \in (\Sigma_a^i \cup \Gamma) \wedge \\&g_i((a,n),e)! \\
		g_i((a,n),e) & \text{if } e \in (\Sigma_a^d \cup \Sigma_o) \wedge\\&g_i((a,n),e)! \wedge\\
		& (a,n) \text{ is not flagged}\\
		\text{undefined} & \text{otherwise}
	\end{array}		
	\right.\]
	\vspace*{-0.5cm}
\end{algorithmic}
\end{algorithm}
Bounded attacks include features from both unbounded and interruptible attacks.
$E$-states that have reached the maximum allowed modification behave as $E$-states in the interruptible case, in other words, they must satisfy the race-free condition. 
On the contrary, $E$-states that have not reached the maximum allowed editions are similar to $E$-states in the unbounded scenario. 
Consequently, Step $2.1$ and Step $2.2$ are a combination of the corresponding steps in the previous cases.
A similar structure as the ISDA and the USDA can be introduced, however we omit such definition given that it would follow the same steps as in the previous cases. 
The same comment holds for the lemmas and the theorems introduced in the previous cases. 
The details are left for the readers to work out.

%
%
%
%
%
\section{Conclusion} \label{sec:conclusion} 
We have considered the supervisory layer of feedback control systems, where sensor readings may be compromised by an attacker in the form of insertions and deletions.
In this context, we have formulated the problem of synthesizing stealthy sensor deception attacks, that can cause damage to the system without detection by an existing supervisor. 
We defined the attacker as a nondeterministic edit function that reacts to the plant's output and its previous editions in a way that guarantees stealthiness of its attack. 
We introduced three different types of attacks, based on the interaction between the system and the attacker. 

Our solution procedure is game-based and relies on the construction of a discrete structure called the AIDA, which is used to solve the synthesis problem for each attack scenario. 
The AIDA captures the game between the environment (i.e., system and attacker) and the given supervisor. 
It embeds all valid actions of the attacker. 
Based on the AIDA, we specified a pruning procedure for each attack type, thereby constructing stealthy structures denoted as the ISDA and the USDA. 
Based on each type of stealthy structure, we can synthesize, if it exists, an attack function that leads the system to unsafe critical states without detection, for the corresponding attack scenario. 

In the future, we plan to investigate how to modify a supervisor that is susceptible to stealthy deception attacks. 
We also plan to study the problem of directly designing supervisors that enforce safety and liveness specifications and at the same time are robust to deception attacks.
\vspace*{-0.5cm}
\bibliographystyle{aut}        
\bibliography{IEEEbib}           

@article{Weerakkody:2019,
year = {2019},
volume = {7},
journal = {Foundations and Trends in Systems and Control},
title = {Resilient Control in Cyber-Physical Systems: Countering Uncertainty, Constraints, and Adversarial Behavior},
issn = {2325-6818},
number = {1-2},
pages = {1-252},
author = {Sean Weerakkody and Omur Ozel and Yilin Mo and Bruno Sinopoli}
}

@article{Meira-Goes:2018,
title = "Demonstration of Indoor Location Privacy Enforcement using Obfuscation",
journal = "14th IFAC Workshop on Discrete Event Systems WODES 2018",
volume = "51",
number = "7",
pages = "145 - 151",
year = "2018",
author = "R\^omulo {Meira-G\'oes} and Blake C. Rawlings and Nicholas Recker and Gregory Willett and St\'ephane Lafortune",
}

@Article{Wu:2018,
author="Wu, Yi-Chin
and Raman, Vasumathi
and Rawlings, Blake C.
and Lafortune, St{\'e}phane
and Seshia, Sanjit A.",
title="Synthesis of Obfuscation Policies to Ensure Privacy and Utility",
journal="Journal of Automated Reasoning",
year="2018",
month="Jan",
day="01",
volume="60",
number="1",
pages="107--131",
}

@article{Carvalho:2018,
title = "Detection and mitigation of classes of attacks in supervisory control systems",
journal = "Automatica",
volume = "97",
pages = "121 - 133",
year = "2018",
author = "Lilian Kawakami Carvalho and Yi-Chin Wu and Raymond Kwong and St\'ephane Lafortune",
}

@article{Su:2018,
title = "Supervisor synthesis to thwart cyber attack with bounded sensor reading alterations",
journal = "Automatica",
volume = "94",
pages = "35-44",
year = "2018",
author = "Rong Su",
}

@INPROCEEDINGS{Goes:2017, 
author={R. {Meira-G\'oes} and E. Kang and R. Kwong and S. Lafortune}, 
booktitle={2017 IEEE 56th Annual Conference on Decision and Control (CDC)}, 
title={Stealthy deception attacks for cyber-physical systems}, 
year={2017}, 
volume={}, 
number={}, 
pages={4224-4230}, 
ISSN={}, 
month={Dec},
}

@article{Wakaiki:2017,
author="Wakaiki, Masashi
and Tabuada, Paulo
and Hespanha, Jo{\~a}o P.",
title="Supervisory Control of Discrete-Event Systems Under Attacks",
journal="Dynamic Games and Applications",
year="2018",
month="Sep",
day="08",  
  }

@ARTICLE{Yin:2016c, 
author={X. Yin}, 
journal={IEEE Transactions on Automatic Control}, 
title={Supervisor Synthesis for Mealy Automata with Output Functions: A Model Transformation Approach}, 
year={2017}, 
volume={62}, 
number={5}, 
pages={2576-2581},
month={},}

@ARTICLE{Yin:2016a, 
author={X. Yin and S. Lafortune}, 
journal={IEEE Transactions on Automatic Control}, 
title={A Uniform Approach for Synthesizing Property-Enforcing Supervisors for Partially-Observed Discrete-Event Systems}, 
year={2016}, 
volume={61}, 
number={8}, 
pages={2140-2154},  
ISSN={0018-9286}, 
month={Aug},}

@ARTICLE{Yin:2016b, 
author={X. Yin and S. Lafortune}, 
journal={IEEE Transactions on Automatic Control}, 
title={Synthesis of Maximally-Permissive Supervisors for the Range Control Problem}, 
year={2017}, 
volume={62}, 
number={8}, 
pages={3914-3929},}

@article{Kerns:2014,
 author = {Kerns, Andrew J. and Shepard, Daniel P. and Bhatti, Jahshan A. and Humphreys, Todd E.},
 title = {Unmanned Aircraft Capture and Control Via {GPS} Spoofing},
 journal = {J. Field Robot.},
 issue_date = {July 2014},
 volume = {31},
 number = {4},
 month = jul,
 year = {2014},
 issn = {1556-4959},
 pages = {617--636},
 numpages = {20},
 acmid = {2885406},
 publisher = {John Wiley and Sons Ltd.},
 address = {Chichester, UK}
}

@INPROCEEDINGS{Alves:2014,
author = "Marcos Vin\'icius S. Alves and Joao Carlos Basilio and Antonio Eduardo C. da Cunha and Lilian Kawakami Carvalho and Marcos Vicente Moreira",
booktitle = "12th IFAC International Workshop on Discrete Event Systems",
title = "Robust Supervisory Control Against Intermittent Loss of Observations",
pages = "294 - 299",
year = "2014",
month={May}}

@article{Lin:2014,
author = {Feng Lin},
title = {Control of Networked Discrete Event Systems: Dealing with Communication Delays and Losses},
journal = {SIAM Journal on Control and Optimization},
volume = {52},
number = {2},
pages = {1276-1298},
year = {2014},
}

@ARTICLE{Amin:2013,
author={S. {Amin} and X. {Litrico} and S. {Sastry} and A. M. {Bayen}},
journal={IEEE Transactions on Control Systems Technology},
title={Cyber Security of Water SCADA Systems - {Part I}: Analysis and Experimentation of Stealthy Deception Attacks},
year={2013},
volume={21},
number={5},
pages={1963-1970},
month={Sep.},}

@INPROCEEDINGS{Rohloff:2012,
author = "Kurt Rohloff",
booktitle = "11th IFAC International Workshop on Discrete Event Systems",
title = "Bounded Sensor Failure Tolerant Supervisory Control",
pages = "272 - 277",
year = "2012",
month={October}}

@Article{Cassez:2012,
author="Cassez, Franck
and Dubreil, J{\'e}r{\'e}my
and Marchand, Herv{\'e}",
title="Synthesis of opaque systems with static and dynamic masks",
journal="Formal Methods in System Design",
year="2012",
volume="40",
number="1",
pages="88--115",
}

@inproceedings{Teixeira:2012,
 author = {Teixeira, Andr{\'e} and P{\'e}rez, Daniel and Sandberg, Henrik and Johansson, Karl Henrik},
 title = {Attack Models and Scenarios for Networked Control Systems},
 booktitle = {Proceedings of the 1st International Conference on High Confidence Networked Systems},
 series = {HiCoNS '12},
 year = {2012},
 isbn = {978-1-4503-1263-9},
 location = {Beijing, China},
 pages = {55--64},
 numpages = {10},
 acmid = {2185515},
 publisher = {ACM},
 address = {New York, NY, USA}
}

@article{Farwell:2011,
author = { James   P.   Farwell  and  Rafal   Rohozinski },
title = {Stuxnet and the Future of Cyber War},
journal = {Survival},
volume = {53},
number = {1},
pages = {23-40},
year = {2011},
}

@inproceedings{Checkoway:2011,
 author = {Checkoway, Stephen and McCoy, Damon and Kantor, Brian and Anderson, Danny and Shacham, Hovav and Savage, Stefan and Koscher, Karl and Czeskis, Alexei and Roesner, Franziska and Kohno, Tadayoshi},
 title = {Comprehensive Experimental Analyses of Automotive Attack Surfaces},
 booktitle = {Proceedings of the 20th USENIX Conference on Security},
 series = {SEC'11},
 year = {2011},
 location = {San Francisco, CA},
 pages = {6--6},
 numpages = {1},
  acmid = {2028073},
 publisher = {USENIX Association},
 address = {Berkeley, CA, USA}
}

@article{Lin:2011,
 author = {Lin, Feng},
 title = {Opacity of Discrete Event Systems and Its Applications},
 journal = {Automatica},
 issue_date = {March, 2011},
 volume = {47},
 number = {3},
 month = mar,
 year = {2011},
 issn = {0005-1098},
 pages = {496--503}
}

@article{Paoli:2011,
 author = {Paoli, Andrea and Sartini, Matteo and Lafortune, St{\'e}phane},
 title = {Active Fault Tolerant Control of Discrete Event Systems Using Online Diagnostics},
 journal = {Automatica},
 issue_date = {April, 2011},
 volume = {47},
 number = {4},
 month = apr,
 year = {2011},
 issn = {0005-1098},
 pages = {639--649},
 numpages = {11},
 acmid = {1963778},
 publisher = {Pergamon Press, Inc.},
 address = {Tarrytown, NY, USA}
}

@INPROCEEDINGS{Xu:2009, 
author={S. Xu and R. Kumar}, 
booktitle={2009 IEEE International Conference on Automation Science and Engineering}, 
title={Discrete event control under nondeterministic partial observation}, 
year={2009}, 
pages={127-132}, 
month={Aug},
}

@INPROCEEDINGS{Cardenas:2008, 
author={A. A. Cardenas and S. Amin and S. Sastry}, 
booktitle={2008 The 28th International Conference on Distributed Computing Systems Workshops}, 
title={Secure Control: Towards Survivable Cyber-Physical Systems}, 
year={2008}, 
pages={495-500},
ISSN={1545-0678}, 
month={June},
}

@book{Lafortune:2008,
 author = {Cassandras, Christos G. and Lafortune, St\'ephane},
 title = {Introduction to Discrete Event Systems},
 year = {2008},
 isbn = {0387333320},
 publisher = {Springer-Verlag New York, Inc.},
 address = {Secaucus, NJ, USA}
}

@INPROCEEDINGS{Thorsley:2006, 
author={D. Thorsley and D. Teneketzis}, 
booktitle={Proceedings of the 45th IEEE Conference on Decision and Control}, 
title={Intrusion Detection in Controlled Discrete Event Systems}, 
year={2006}, 
pages={6047-6054}, 
ISSN={0191-2216}, 
month={Dec},}

@INPROCEEDINGS{Saboori:2007, 
author={A. Saboori and C. N. Hadjicostis}, 
booktitle={46th IEEE Conference on Decision and Control}, 
title={Notions of security and opacity in discrete event systems}, 
year={2007}, 
pages={5056-5061}, 
month={Dec},}

@article{Paoli:2005,
 author = {Paoli, Andrea and Lafortune, St{\'e}phane},
 title = {Safe Diagnosability for Fault-tolerant Supervision of Discrete-event Systems},
 journal = {Automatica},
 issue_date = {August, 2005},
 volume = {41},
 number = {8},
 month = aug,
 year = {2005},
 pages = {1335--1347},
 numpages = {13},
 publisher = {Pergamon Press, Inc.},
 address = {Tarrytown, NY, USA}
}

@INPROCEEDINGS{Golaszewski:1987, 
author={C. H. Golaszewski and P. J. Ramadge}, 
booktitle={26th IEEE Conference on Decision and Control}, 
title={Control of discrete event processes with forced events}, 
year={1987}, 
volume={26}, 
number={}, 
pages={247-251},
month={Dec},}

@article{Ramadge:1987,
 author = {Ramadge, P. J. and Wonham, W. M.},
 title = {Supervisory Control of a Class of Discrete Event Processes},
 journal = {SIAM Journal on Control and Optimization},
 issue_date = {Jan. 1987},
 volume = {25},
 number = {1},
 month = jan,
 year = {1987},
 issn = {0363-0129},
 pages = {206--230},
 numpages = {25},
 acmid = {35482},
 publisher = {Society for Industrial and Applied Mathematics},
 address = {Philadelphia, PA, USA}
}
%
\appendix

\section{Proofs}

\textbf{Proof of Lemma~\ref{lemma:supervisor_estimate}}

\begin{proof}
We prove the result by induction on the length of $s_A$.

Let $|s_A| = n$.
Let $y_0$ be defined as usual, and 
$z_i = h_{SE}(y_i,S^d_A(s_A^i))$  

for $i \in \mathbb{N}^n$ and $y_{i+1} = h_{ES}(z_i,e^{i+1})$ for $i \in \mathbb{N}^{n-1}$.

Induction Basis: $s_A = \epsilon$
\begin{equation*}
IE(\epsilon) = h_{SE}(y_0,S_P(\epsilon)) 
\end{equation*}
We recall that the supervisor state does not change in the IDA from S-states to E-states since it did not receive any observable event, thus   
$y_k \rightarrow z_k \Rightarrow I_S(y_k) = I_S(z_k)$, then
\begin{equation*}
I_S(IE(\epsilon)) = q_0 = \tilde{\mu}(q_0,\epsilon)
\end{equation*}

Induction hypothesis: Assume that $I_S(IE(s_A^i))=\tilde{\mu}(q_0,P_e^S(s^i_A))$ holds for $i\in \mathbb{N}^k$, $k<n$. 

Induction step: At $k+1$ we have
\begin{equation*}
y_{k+1} = h_{ES}(z_k,e^{k+1})
\end{equation*}
And we know that 
\begin{equation*}
I_S(z_k) = \tilde{\mu}(q_0,P_e^S(s_A^k))
\end{equation*}
\begin{equation*}
I_S(y_{k+1}) = \tilde{\mu}(I_S(z_k),P_e^S(e^{k+1})) = \tilde{\mu}(q_0,P_e^S(s_A^{k+1}))
\end{equation*}
Given that $I_S(y_{k+1}) = I_S(z_{k+1})$, then  $I_S(z_{k+1}) = \tilde{\mu}(q_0,P_e^S(s_A^{k+1}))$. 
Consequently, $I_S(IE(s_A^{k+1})) = \tilde{\mu}(q_0,\allowbreak P_e^S(s_A^{k+1}))$.
\end{proof}

\textbf{Proof of Theorem~\ref{theo:AIDA}}
\begin{proof} 
Conditions 1 and 2 of Definition~\ref{def:AIDA} follow directly by the construction of the Algorithm \ref{algo:construction} (lines $14$, $29$, $30$). Thus, we only need to prove that the IDA returned by the algorithm is the largest one. We show it by contradiction.

Assume that $A$ is the IDA returned by the algorithm; however assume that $\exists A^*$ that satisfies conditions 1 and 2, 
where $A\sqsubset A^*$. 
It means that $Q^A_S \subseteq Q^{A^*}_S$ or $Q^A_E \subseteq Q^{A^*}_E$. 
If $A$ and $A^*$ have the same states, then either $A = A^*$ or $A^*$ has more transitions than $A$. 
The first case is a contradiction of our arguments. 
The second case implies that some transitions were not included in $A$, which is also a contradiction. 
Algorithm \ref{algo:construction} applies an exhaustive BFS therefore it cannot leave it out any transition  if all states where covered.
Thus, it is the case that $A^*$ has more states than $A$.
Let us start with $A^*$ having one additional $E$-state namely $z$, then $Q^A_S = Q^{A^*}_S$. 
Therefore, $\exists y \in Q^A_S$ such that $y \rightarrow z$, which means that $I_S(y)\neq$ \emph{dead}. 
Therefore, Algorithm \ref{algo:construction} will not converge to $A$, contradicting our assumption. 
The same reasoning can be used for the case of one more $S$-state or when both sets are larger. 
\end{proof}

\textbf{Proof of Lemma~\ref{lem:f_AinISDA}}

\begin{proof}
By contradiction, assume that we have an interruptible $f_A$ that is admissible and stealthy, but is not embedded in the ISDA. 
There exists a string $s \in \mathcal{L}(S_A/G)$, $s_a \in \hat{f}_A(P_o(s))$ where $IE(s_a)$ is not defined in the ISDA. 
For simplicity and without loss of generality, assume that $s_a = t_ae$ where $e\in \Sigma_o\cup\Sigma_a^e$, and $IE(t_a)$ is defined but $IE(s_a)$ is not defined.
There are two reasons why $IE(s_a)$ is not defined.
\begin{enumerate}
\item $IE(s_a)$ is not defined in the AIDA. 
We have that $z = IE(t_a)$ in the AIDA, however $IE(z,e)$ is not defined. 
Based on the construction of the AIDA, at state $z$ the transition function is exhaustively constructed given $\Gamma_{\tilde{R}}(I_S(z))$. 
Therefore, if $IE(z,e)$ is not defined, then $\mathcal{M}_e(e)\notin \Gamma_{\tilde{R}}(I_S(z))$.
Since $\tilde{R}$ is admissible, $\mathcal{M}_e(e) \in \Sigma_c\cap\Sigma_o$ and $e\in \Sigma_a^i$.
As consequence of Lemma~\ref{lemma:supervisor_estimate} $P_e^S(s_a)\not\in P_o(\lang(S_P/G))$, which trivially violates stealthiness.
This contradicts our assumption of stealthy $f_A$.
Note that, this case is different than reaching the dead state.
In this case, the attacker inserts an event that is not allowed by the supervisor.
%
\item 
$IE(s_a)$ is defined in the AIDA but it was pruned by Algorithm \ref{algo:BSCP}. Algorithm \ref{algo:BSCP} returns the ``supremal controllable sublanguage'' of the AIDA, i.e., it is maximally permissive, under the race-free and controllability conditions. 
It removes all sequences that are non-stealthy or that uncontrollably lead to non-stealthiness.
Similarly for the race-free condition.
Moreover, one cannot define an interruptible attack decision at a state that is not race free.
Finally, the definition of the set of controllable events guarantees admissibility.
Thus, overall, $s_a$ must lead to a non-stealthy, non-interruptible, or inadmissible strategy, which makes the function $f_A$ also either non-stealthy, non-interruptible, or inadmissible, a contradiction.
\end{enumerate} 
This completes the proof.
\end{proof}

\textbf{Proof of Lemma~\ref{lem:f_AsyntesisISDA}}
\begin{proof}
The proof follows directly by the construction of the ISDA.
\end{proof}

\textbf{Proof of Theorem~\ref{thm:all}}
\begin{proof}
The proof follows from Lemmas \ref{lem:f_AinISDA} and \ref{lem:f_AsyntesisISDA}.
\end{proof}

\textbf{Proof of Theorem~\ref{thm:main}}
\begin{proof}
The proof follows from Theorem \ref{thm:all}.
\end{proof}

\end{document}